\documentclass[acmlarge]{acmart}
\usepackage{tabularx}
\usepackage{mydef}
\usepackage{dsfont}
\usepackage{xcolor}
\AtBeginDocument{%
  }

\setcopyright{acmlicensed}
\copyrightyear{2025}
\acmYear{2025}
\acmDOI{XXXXXXX.XXXXXXX}

\acmJournal{JACM}
\acmArticle{1}
\acmMonth{1}

\usepackage{amsmath,amsfonts,bm}

\def\eqref#1{equation~\ref{#1}}
\def\1{\bm{1}}

\DeclareMathAlphabet{\mathsfit}{\encodingdefault}{\sfdefault}{m}{sl}
\SetMathAlphabet{\mathsfit}{bold}{\encodingdefault}{\sfdefault}{bx}{n}

\begin{document}

%%
%% The "title" command has an optional parameter,
%% allowing the author to define a "short title" to be used in page headers.
% \title{A Comprehensive Survey of Cold-Start Recommendation: Transition from Contextual Data to Large Language Models\\}

% \title{A Comprehensive Survey of Cold-Start Recommendation in The Era of LLMs}

% \title{Where to Go Next for Cold-Start Recommendation: A Comprehensive Survey from Contextual Data to Large Language Models}

%\title{A Comprehensive Survey on Cold-Start Recommendation: A Roadmap from Contextual Data towards Large Language Models}

% \title{A Roadmap of Cold-Start Recommendation: from Contextual Data towards Large Language Models}

% \title{Cold-Start Recommendation in the Era of Large Language Models: A Roadmap}
%%
\title{Cold-Start Recommendation towards the Era of Large Language Models (LLMs): A Comprehensive Survey and Roadmap}
% \title{From CF to Large Language Model: A Survey of Cold-Start Problem}
%%
%%
%% The "author" command and its associated commands are used to define
%% the authors and their affiliations.
%% Of note is the shared affiliation of the first two authors, and the
%% "authornote" and "authornotemark" commands
%% used to denote shared contribution to the research.
\author{Weizhi Zhang}
\authornote{Both authors contributed equally to this research.}
\email{wzhan42@uic.edu}
\orcid{0000-0003-4067-7588}
\affiliation{%
  \institution{University of Illinois Chicago}
  % \city{Chicago}
  % \state{Illinois}
  \country{USA}
}

\author{Yuanchen Bei}
\authornotemark[1]
\email{yuanchenbei@zju.edu.cn}
\orcid{0000-0003-2834-2873}
\affiliation{%
  \institution{Zhejiang University}
  % \city{Hangzhou}
  \country{China}
}

\author{Liangwei Yang}
\email{lyang84@uic.edu}
\orcid{0000-0001-5660-766X}
\author{Henry Peng Zou}
\email{pzou3@uic.edu}
\orcid{0009-0003-5259-4998}
\affiliation{%
  \institution{University of Illinois Chicago}
  % \city{Chicago}
  % \state{Illinois}
  \country{USA}
}

\author{Peilin Zhou}
\email{pzhou460@connect.hkust-gz.edu.cn}
\orcid{}
\affiliation{%
  \institution{Hong Kong University of Science and Technology (Guangzhou)}
  % \city{Guangzhou}
  \country{China}
}

\author{Aiwei Liu}
\email{liuaw20@mails.tsinghua.edu.cn}
\orcid{0000-0002-4965-8263}
\author{Yinghui Li}
\email{liyinghu20@mails.tsinghua.edu.cn}
\orcid{}
\affiliation{%
  \institution{Tsinghua University}
  % \city{Beijing}
  % \state{}
  \country{China}
}

\author{Hao Chen}
\email{sundaychenhao@gmail.com}
\orcid{0000-0001-6816-5344}
\affiliation{%
  \institution{The Hong Kong Polytechnic University}
  % \city{HongKong SAR}
  \country{China}
}

\author{Jianling Wang}
\email{jianlingw@google.com}
\orcid{}
\affiliation{%
  \institution{Google Deepmind}
  % \city{Mountain View}
  % \state{California}
  \country{USA}
}

\author{Yu Wang}
\email{yuw@Netflix.com}
\orcid{}
\affiliation{%
  \institution{Netflix}
  % \city{Los Gatos}
  % \state{California}
  \country{USA}
}

\author{Feiran Huang}
\email{huangfr@jnu.edu.cn}
\orcid{}
\affiliation{%
  \institution{Jinan University}
  % \city{Abu Dhabi}
  \country{China}
}

\author{Sheng Zhou}
\email{zhousheng\_zju@zju.edu.cn}
\orcid{}
\author{Jiajun Bu}
\email{bjj@zju.edu.cn}
\orcid{}
\affiliation{%
  \institution{Zhejiang University}
  % \city{Hangzhou}
  \country{China}
}

\author{Allen Lin}
\email{al001@tamu.edu}
\orcid{}
\author{James Caverlee}
\email{caverlee@tamu.edu}
\orcid{}
\affiliation{%
  \institution{Texas A\&M University}
  % \city{College Station}
  % \state{Texas}
  \country{USA}
}

\author{Fakhri Karray}
\email{Fakhri.Karray@mbzuai.ac.ae}
\orcid{}
\affiliation{%
  \institution{Mohamed Bin Zayed University of Artificial Intelligence}
  % \city{Abu Dhabi}
  \country{UAE}
}

\author{Irwin King}
\email{king@cse.cuhk.edu.hk}
\orcid{0000-0001-8106-6447}
\affiliation{%
  \institution{The Chinese University of Hong Kong}
  % \city{Hong Kong}
  \country{China}
}

\author{Philip S. Yu}
\email{psyu@uic.edu}
\orcid{0000-0002-3491-5968}
\affiliation{%
  \institution{University of Illinois Chicago}
  % \city{Chicago}
  % \state{Illinois}
  \country{USA}}

\renewcommand{\shortauthors}{Weizhi Zhang, Yuanchen Bei, et al.}

%%
%% By default, the full list of authors will be used in the page
%% headers. Often, this list is too long, and will overlap
%% other information printed in the page headers. This command allows
%% the author to define a more concise list
%% of authors' names for this purpose.
% \renewcommand{\shortauthors}{ et al.}

%%
%% The abstract is a short summary of the work to be presented in the
%% article.

\begin{abstract}
Cold-start problem is one of the long-standing challenges in recommender systems, focusing on accurately modeling new or interaction-limited users or items to provide better recommendations. Due to the diversification of internet platforms and the exponential growth of users and items, the importance of cold-start recommendation (CSR) is becoming increasingly evident. At the same time, large language models (LLMs) have achieved tremendous success and possess strong capabilities in modeling user and item information, providing new potential for cold-start recommendations. However, the research community on CSR still lacks a comprehensive review and reflection in this field. Based on this, in this paper, we stand in the context of the era of large language models and provide a comprehensive review and discussion on the roadmap, related literature, and future directions of CSR. Specifically, we have conducted an exploration of the development path of how existing CSR utilizes information, from content features, graph relations, and domain information, to the world knowledge possessed by large language models, aiming to provide new insights for both the research and industrial communities on CSR.
Related resources of cold-start recommendations are collected and continuously updated for the community in \textcolor{blue}{\url{https://github.com/YuanchenBei/Awesome-Cold-Start-Recommendation}}.

\end{abstract}

%%
%% The code below is generated by the tool at http://dl.acm.org/ccs.cfm.
%% Please copy and paste the code instead of the example below.
%%

%%
%% Keywords. The author(s) should pick words that accurately describe
%% the work being presented. Separate the keywords with commas.

% \received{20 February 2007}
% \received[revised]{12 March 2009}
% \received[accepted]{5 June 2009}

%%
%% This command processes the author and affiliation and title
%% information and builds the first part of the formatted document.
\maketitle

\setcounter{tocdepth}{3}
\textcolor{blue}{\parskip 0.5pt \tableofcontents}
%\small 

\clearpage

\section{INTRODUCTION}

\begin{figure}[b]
     \centering
     \includegraphics[width=1\linewidth, trim=0cm 0cm 0cm 0cm,clip]{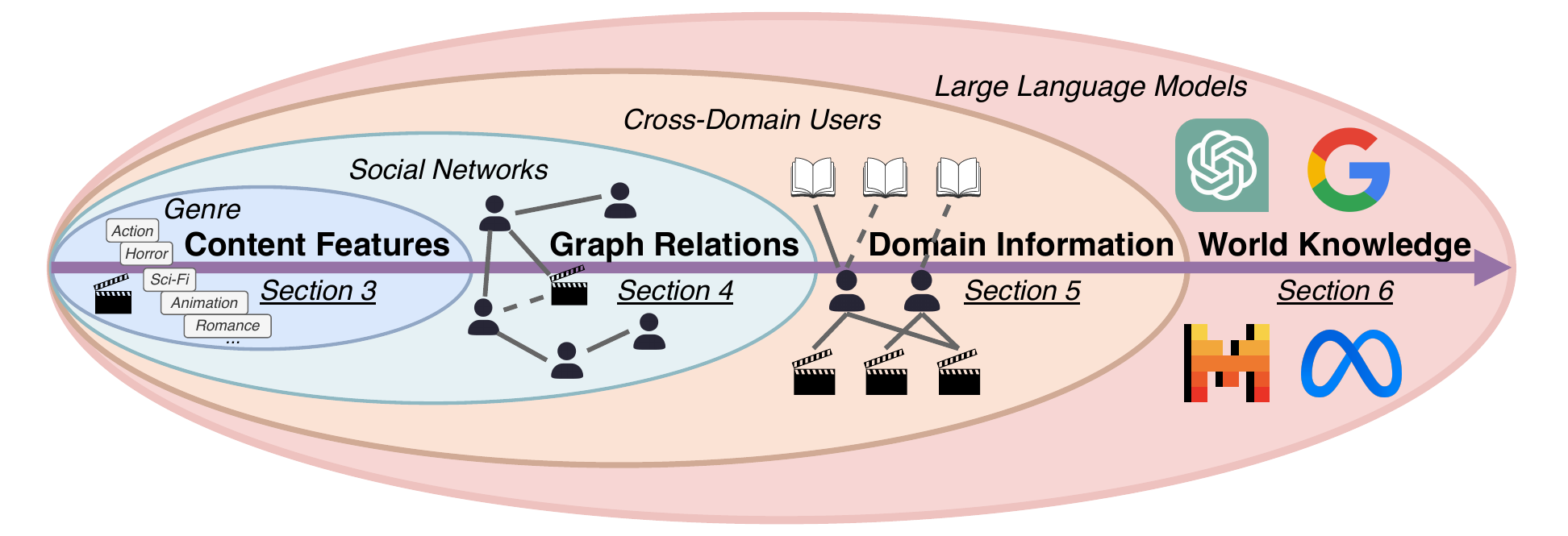}
     \caption{Illustrations of the knowledge scope discussed in this survey.}
     \label{fig:scope}
\end{figure}

In the rapidly evolving landscape of the digital information era, recommender systems (RecSys) have become indispensable tools for helping users discover relevant content and items amidst overwhelming information and choices~\cite{zhang2019deep,wu2022graph,bobadilla2013recommender}. Despite their widespread deployment, RecSys face persistent challenges, particularly in "cold-start" scenarios, where limited or no historical interaction data is available for new users or items. Specifically, in real-world scenarios, the cold-start problem could be the introduction of new items, the onboarding of new users, or new platforms with inherently sparse interaction data. 
Addressing the cold-start problem is not just technically necessary for performance metrics but also critical for advancing the effectiveness and sustainability of recommender systems. First and foremost, solving this issue ensures that new users and items are fairly represented, mitigating biases that arise from the reliance on historical data. This improvement fosters diversity and fairness in recommendations, promoting diverse content exposure by preventing new items from being overlooked~\cite{lin2022quantifying, zhu2021fairness}. 
Furthermore, tackling the cold-start challenge brings the platform brand value and user retention. In a crowded and fast-moving digital landscape, delivering immediate and relevant recommendations to new users can differentiate a platform from its competitors. Personalized recommendations from the outset help engage new users, preventing them from leaving due to irrelevant or absent suggestions. This creates a strong initial impression and fosters loyalty.
For platforms, this translates into higher engagement, improved retention rates, and the ability to succeed in dynamic markets. 
Finally, effectively addressing cold-start scenarios ensures scalability and growth. As platforms expand with new users and content, effective integration of the continual influx of these entities keeps recommendation engines dynamic and relevant. This adaptability supports long-term sustainability in a rapidly changing environment.
Given these motivations, the cold-start problem has driven the exploration of innovative approaches that leverage diverse external knowledge sources. By incorporating information beyond traditional user-item interactions~\cite{he2017neural,wang2019neural}, such as content features~\cite{volkovs2017dropoutnet}, social information~\cite{du2022socially}, or pretrained LLM knowledge~\cite{liu2023chatgpt}, these methods enrich the representation and modeling of cold-start entities, enabling recommender systems to perform effectively even under sparse data conditions. As such, solving the cold-start problem is not merely a technical challenge—it is a strategic necessity for building fair, engaging, and sustainable recommendation platforms in an ever-changing digital landscape.

\begin{figure}[htbp]
    \centering
    % First figure
    \begin{subfigure}[b]{.5\linewidth}
        \centering
        \includegraphics[width=\textwidth]{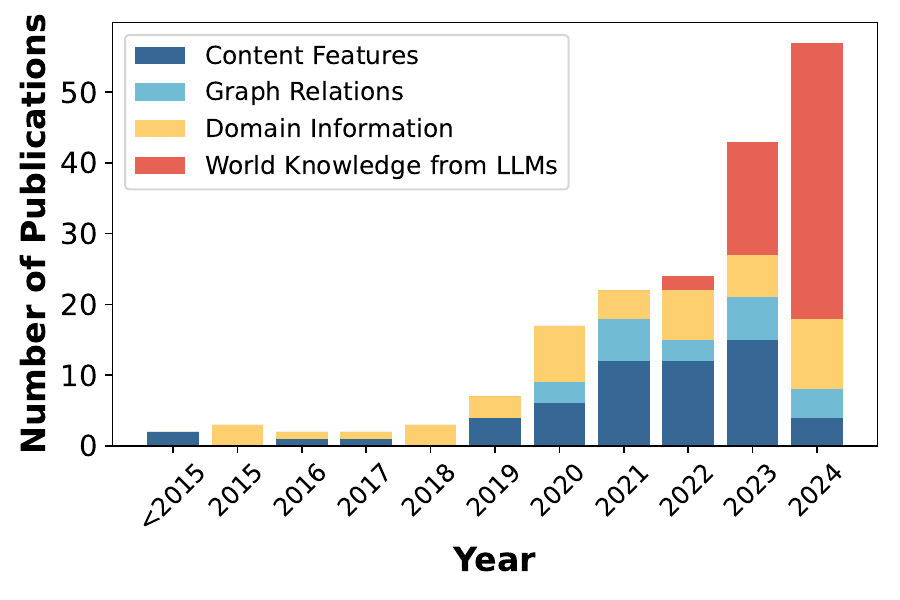}  
        \caption{Number of publications in recent years.}
        \label{fig:trend}
    \end{subfigure}
    \hfill
    % Second figure
    \begin{subfigure}[b]{.4\linewidth}
        \centering
        \includegraphics[width=\textwidth]{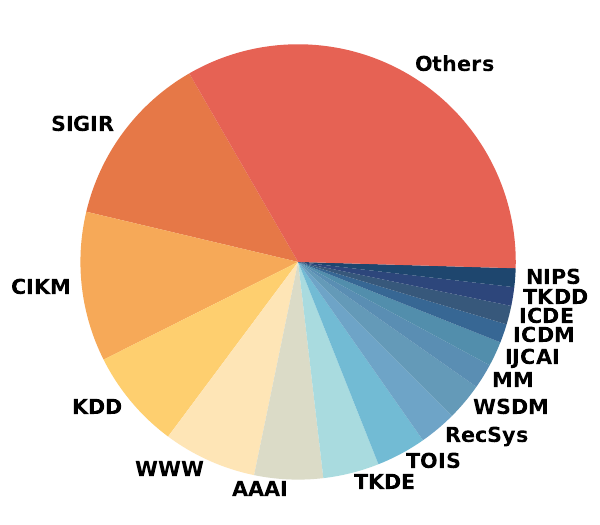}  
        \caption{Venues of publications}
        \label{fig:pie}
    \end{subfigure}
    \caption{Statistical information of current cold-start recommendation publications.}
    \label{fig:stat}
\end{figure}

Early cold-start attempts adopt content-based approaches~\cite{van2013deep,lops2011content} and focus on categorical textual features, such as item genres, item titles, and user profiles, which play a crucial role in representing cold entities. 
Then, with the advances of graph mining techniques \cite{kipf2022semi,wu2020comprehensive,xu2018powerful}, high-order relations derived from graph structures, such as user-item interaction graphs~\cite{he2020lightgcn,chen2024macro,zhang2024mixed,zhang2024we}, knowledge graphs~\cite{cao2019unifying,wang2019kgat}, and social networks~\cite{yang2021consisrec,sharma2024survey}, become another critical component to enhance the CSR.
Concurrently, instead of mining the graph relations among nodes, some researchers resort to the connections among different domains~\cite{khan2017cross,zang2022survey}. In particular, cold-start and data sparsity issues in the target domain can be alleviated by transferring knowledge from other domains where richer interaction data is available. Cross-domain recommendation techniques exploit overlapping user bases, shared attributes, or aligned item categories to improve performance in CSR. 
In recent years, the rise of large language models (LLMs), such as GPT~\cite{achiam2023gpt}, LLaMa~\cite{dubey2024llama}, and T5~\cite{raffel2020exploring} has revolutionized natural language processing, demonstrating exceptional capabilities in understanding and generating human-like text based on vast amounts of pre-trained data~\cite{min2023recent,li2024pre}. These advancements have inspired a paradigm shift in recommender system research, leveraging the comprehensive contextual understanding of LLMs to enhance cold-start recommendation performance. By utilizing the pre-trained world knowledge of LLMs, researchers have started exploring novel strategies for modeling and representing cold users and items in a more semantically rich and context-aware manner.  Figure~\ref{fig:trend} illustrates this evolving trend, highlighting the shift in cold-start recommendation research from traditional content-based methods to LLM-driven strategies with progressively expanding knowledge scope (Figure~\ref{fig:scope}).

% Although there have been several survey paperss covering the CSRecSys, they lack a systematic review, which is critical for further/deeper development.
This survey paper aims to provide an extensive review of state-of-the-art techniques and frameworks in the cold-start recommendation, with a special outlook towards the era of LLMs with expanding knowledge scope as illustrated in Figure~\ref{fig:scope}. Particular focuses are placed on works published in top-tier conferences and journals, as in Figure~\ref{fig:pie}. Based on the collection, we categorize the existing works into four knowledge scopes considering the scale of external knowledge sources: Content Features, Graph Relations, Domain Information, and World Knowledge from the Large Language Models.
By systematically categorizing and analyzing these approaches, our survey aims to present a comprehensive understanding of the landscape and propose a roadmap for future research. We emphasize the transformative potential of integrating LLMs into cold-start recommendations and outline the opportunities and challenges that lie ahead in this burgeoning field.

\subsection{Related Work}
A comparison between our survey and the previous surveys is shown in Table~\ref{tab:survey}. All of the existing surveys, which cover cold-start recommendation articles, only focus on partial knowledge scopes or limited aspects of the CSR problem. The earliest surveys~\cite{gope2017survey} and~\cite{camacho2018social} partially covered single knowledge scope without defining specific cold-start issues. Later surveys from IDCTIT~\cite{sethi2021cold} and Applied Sciences~\cite{abdullah2021eliciting} began incorporating graph relation and domain information, and being the first to explicitly define system cold-start and user cold-start issues, covering more relevant papers through 2021.
More recent surveys such as JIIS~\cite{panda2022approaches} and IEEE Access~\cite{yuan2023user} have expanded the scope and number of covered papers, with \cite{yuan2023user} particularly focusing on user cold-start problems.  
In all, no existing survey paper in the literature fully covers all four aspects (Features, Graph, Domain, and LLMs) while addressing multiple cold-start issues. In this work, we aim to fill this gap by providing a comprehensive and systematic survey covering 220 papers through December 2024, clearly defining 9 distinct cold-start issues and incorporating analysis across knowledge scopes from features, graphs, domains, and LLMs.

\begin{table*}[t]
    \centering
    \resizebox{\textwidth}{!}{
    \begin{threeparttable}
    \small
    \caption{Comparison with existing surveys. For each survey, we summarize the knowledge scope covered in their collected relevant papers, the corresponding statistics, and the specific types of cold-start issues defined and discussed in the survey.}
    \newcolumntype{H}{>{\setbox0=\hbox\bgroup}c<{\egroup}@{}}
    \begin{tabular}{c|c|cccc|cc|c}
        \toprule
        \multirow{2}{*}{\textbf{Surveys}} & \multirow{2}{*}{\textbf{Venues}} &
        \multicolumn{4}{c|}{\textbf{Knowledeg Scope}} & 
        \multicolumn{2}{c|}{\textbf{Cold-Start RecSys Papers}} & 
        \multirow{2}{*}{\textbf{Defined/Focused Issue}} \\
        & & \textbf{Features} & \textbf{Graph} & \textbf{Domain} & \textbf{LLMs} & \textbf{\# Papers} & \textbf{Latest Year} &  \\

        \midrule
        \cite{gope2017survey} & ICCCA* & \dashcheckmark & & & &  $8$ & 2017 & - \\ %International Conference on Computing, Communication and Automation 
        \cite{camacho2018social} & IPM*  & & \dashcheckmark & & & $14$  & 2018 & - \\ %Information Processing and Management
        \cite{sethi2021cold} & IDCTIT*  & \dashcheckmark & \dashcheckmark & & & $18$ & 2020 & System Cold-Start  \\ % Intelligent Data Communication Technologies and Internet of Things
        \cite{abdullah2021eliciting} & Applied Sciences & \dashcheckmark &  & \dashcheckmark &  & $50$ & 2021 & User Cold-Start\\ %Applied Sciences        
        \cite{panda2022approaches} & JIIS*  & \dashcheckmark & \dashcheckmark & \dashcheckmark &  & $91$ & 2022 & - \\ %Journal of Intelligent Information Systems %66
        \cite{yuan2023user} & IEEE Access & \dashcheckmark & \dashcheckmark & \dashcheckmark &  & $45$ & 2023 & User Cold-Start \\ %56
        \midrule 
        
        \multicolumn{2}{c|}{\textbf{Ours}} & \cmark & \cmark & \cmark & \cmark & \textbf{220} & \textbf{Dec, 2024} & {\color{blue}Cold-Start}  \\
        \bottomrule
    \end{tabular}\label{tab:survey}
    \begin{tablenotes}
        \footnotesize
        \item \cmark: fully covered, \dashcheckmark: partially covered. {\color{blue}Cold-Start}: covers 9 sub cold-start tasks as in Table~\ref{tab:CSR_problem} and Figure~\ref{fig:CSR_problem}.
        \item Latest year: the latest publication year of a relevant paper included in the survey.
        \item ICCCA*: International Conference on Computing, Communication and Automation; IPM*: Information Processing and Management.
        \item ICDTIT*: Intelligent Data Communication Technologies and Internet of Things; JIIS*: Journal of Intelligent Information Systems.
    \end{tablenotes}
    \end{threeparttable}
    }
\end{table*}

% \iffalse
\tikzstyle{leaf0}=[draw=black, %边框
    rounded corners,minimum height=1.2em,
    edge=black!10, 
    text opacity=1, align=center,
    fill opacity=.3,  text=black,font=\scriptsize, %0.3
    inner xsep=2pt, inner ysep=4.2pt,
    ]
\tikzstyle{leaf1}=[draw=black, %边框
    rounded corners,minimum height=1.2em,
    edge=black!10, 
    text opacity=1, align=center,
    fill opacity=.3,  text=black,font=\scriptsize, %0.3
    inner xsep=2pt, inner ysep=4.2pt,
    ]
\tikzstyle{leaf2}=[draw=black, %边框
    rounded corners,minimum height=1.2em,
    edge=black!10, 
    text opacity=1, align=center,
    fill opacity=.5,  text=black,font=\scriptsize,
    inner xsep=2pt, inner ysep=4.5pt,
    ]
\tikzstyle{leaf3}=[draw=black, %边框
    rounded corners,minimum height=1.2em,
    edge=black!10, 
    text opacity=1, align=center,
    fill opacity=0.8,  text=black,font=\scriptsize, %0.8
    inner xsep=2pt, inner ysep=4.5pt,
    ]
\tikzstyle{leaf4}=[draw=black, %边框
    rounded corners,minimum height=1.2em,
    text width=4.4em, 
    edge=black!10, 
    text opacity=1, align=center,
    fill opacity=1,  text=black,font=\scriptsize,
    inner xsep=2pt, inner ysep=4.5pt,
    ]
\begin{figure}[htbp]
\centering
\begin{forest}
  for tree={
  forked edges,
  grow=east,
  reversed=true,
  anchor=base west,
  parent anchor=east,
  child anchor=west,
  base=middle,
  font=\scriptsize,
  rectangle,
  draw=black, %hiddendraw 所有边框
  edge=black!50, 
  rounded corners,
  minimum width=2em,
  s sep=5pt,
  inner xsep=3pt,
  inner ysep=2pt
  },
  % where level=1{align=center}{},
  % where level=2{text width=6em,font=\scriptsize}{},
  % where level=3{font=\scriptsize}{},
  % where level=4{font=\scriptsize}{},
  % where level=5{font=\scriptsize}{},
  [\textbf{Cold-Start Recommendation (CSR)},rotate=90,anchor=north,edge=black!50,fill=mywhite,fill opacity=1,draw=black,
    %%%
    [\textbf{Content Features} \textcolor{blue}{Section~\ref{sec:content}},leaf4,edge=black!50, fill=mypurple, minimum height=1.2em
        [Data-Incomplete \\ Learning,leaf3,edge=black!50,fill=mypurple, 
            [Robust Co-Training,leaf2,fill=mypurple,
                [Robust Generalization~\cite{volkovs2017dropoutnet,lin2024temporally,li2022transform,zhu2020recommendation,du2020learn}
                {;}
                Autoencoders~\cite{kong2024collaborative,bai2023gorec,zhao2022improving,xu2022alleviating,feng2021zero,li2019zero}, leaf1,fill=mypurple
                ]
            ]
            [Knowledge Alignment,leaf2,fill=mypurple
                [Contrastive Learning~\cite{wang2024preference,xu2024cmclrec,zhou2023contrastive,jiang2023self,wei2021contrastive}
                {;}
                Knowledge Distillation\\~\cite{zhang2023cold,huang2023aligning,zhang2023dual} {;}
                Generative Adversarial Networks~\cite{chen2022generative,alshehri2022generative,zhang2022revisiting}
                ,leaf1,fill=mypurple
                ]
            ]
            [Cold Exploration,leaf2,fill=mypurple
                [Reinforcement Learning~\cite{ji2021reinforcement,chu2023meta,silva2023user}
                {;} 
                Architecture Search~\cite{wu2023coldnas}
                ,leaf1,fill=mypurple
                ]
            ]
            [Feature Similarity Measurement,leaf2,fill=mypurple
                [Multi-Feature Fusion~\cite{jin2023automatic,ma2023cross,tao2022sminet,yuan2016solving,wang2011collaborative}
                {;} 
                Hashing~\cite{hansen2020content,xu2020multi,zhang2020deep,xu2021multi,peng2022binary}
                ,leaf1,fill=mypurple
                ]
            ]
        ]
        [Data-Efficient \\ Learning,leaf3,edge=black!50,fill=mypurple, 
            [Meta-Learning Optimization,leaf2, fill=mypurple
                [Pretraining~\cite{lee2019melu,bharadhwaj2019meta,sun2021form,pang2022pnmta}; Adaptation~\cite{dong2020mamo,pang2022pnmta,feng2021cmml,wang2023preference},leaf1,fill=mypurple
                    % [
                    % MeLU~\cite{lee2019melu};
                    % MetaCS~\cite{bharadhwaj2019meta};
                    % FORM~\cite{sun2021form};
                    % PNMTA~\cite{pang2022pnmta}
                    % ,leaf0,fill=mypurple
                    % ]
                % ]
                % [Adaptation,leaf1,fill=mypurple
                %     [
                %     MAMO~\cite{dong2020mamo};
                %     PNMTA~\cite{pang2022pnmta};
                %     CMML~\cite{feng2021cmml};
                %     PDMA~\cite{wang2023preference}
                %     ,leaf0,fill=mypurple
                %     ]
                ]
            ]
            [Meta-Task Utilization,leaf2,fill=mypurple
                [Task Difference~\cite{wen2023modeling,zhao2023task,kim2023meta}; Task Relevance~\cite{yu2021personalized,lin2021task,yang2022task,wu2023m2eu},leaf1,fill=mypurple
                %     [
                %     WDoF~\cite{wen2023modeling};
                %     TDAS~\cite{zhao2023task};
                %     MELO~\cite{kim2023meta}
                %     ,leaf0,fill=mypurple
                %     ]
                % ]
                % [Task Relevance,leaf1,fill=mypurple
                %     [
                %     PAML~\cite{yu2021personalized};
                %     TaNP~\cite{lin2021task};
                %     TSAML~\cite{yang2022task};
                %     M2EU~\cite{wu2023m2eu}
                %     ,leaf0,fill=mypurple
                %     ]
                ] 
            ]
            [Meta-Embedding Initialization~\cite{pan2019warm, zhu2021learning},leaf2,fill=mypurple
                % [
                % Meta-Embedding~\cite{pan2019warm};
                % MWUF~\cite{zhu2021learning}
                % ,leaf1,fill=mypurple
                % ] 
            ]
            [Sequtial Meta-Learning~\cite{wang2021sequential,zheng2021cold,neupane2022dynamic,pan2022multimodal},leaf2,fill=mypurple
                % [
                % MetaTL~\cite{wang2021sequential};
                % Mecos~\cite{zheng2021cold};
                % Dynamic Meta-Learning~\cite{neupane2022dynamic};
                % MML~\cite{pan2022multimodal}
                % ,leaf1,fill=mypurple
                % ] 
            ]
        ]
    ]
    %%%  
    [\textbf{Graph Relations} \textcolor{blue}{Section~\ref{sec:graph}},leaf4,edge=black!50, fill=myyellow, minimum height=1.2em
        [Interaction Graph  Enhancement,leaf3,edge=black!50, fill=myyellow, minimum height=1.2em,
            [Supplementary Graph Relation~\cite{kim2024content,wang2024mutual,liu2023uncertainty,liu2024fine}; Homophily Network Relation~\cite{wang2021preference,shams2021cluster,ouyang2021learning,du2022socially,sbandi2024mitigating},leaf2,fill=myyellow
                %[CGRC~\cite{kim2024content}; MI-GCN~\cite{wang2024mutual}; UCC~\cite{liu2023uncertainty},leaf1,fill=myyellow
                %]
            ]
            %[Homophily Network Relation~\cite{wang2021preference,shams2021cluster,ouyang2021learning,du2022socially},leaf2,fill=myyellow
                %[PAML~\cite{wang2021preference}; Shams et.al~\cite{shams2021cluster}; GME~\cite{ouyang2021learning}; SDCRec~\cite{du2022socially}  
                %,leaf1,fill=myyellow
                %]
            %]
        ]
        [Graph Relation Extension,leaf3,edge=black!50, fill=myyellow, minimum height=1.2em,
            [Heterogeneous Graph Relation~\cite{liu2020heterogeneous,lu2020meta,wang2021privileged,zheng2021multi,cao2022gift,cai2023user}; Attributed Graph \\Relation~\cite{cao2023multi,wang2024warming}; Knowledge Graph Relation~\cite{du2022metakg,han2024cold,togashi2021alleviating,pan2024meta},leaf2,fill=myyellow
                %[HGNR~\cite{liu2020heterogeneous}; MetaHIN~\cite{lu2020meta}; PGD~\cite{wang2021privileged}; MvDGAE~\cite{zheng2021multi}; GIFT~\cite{cao2022gift}; IHGNN~\cite{cai2023user}
                %,leaf1,fill=myyellow
                %]
            ]
            %[Attributed Graph Relation~\cite{cao2023multi,wang2024warming},leaf2,fill=myyellow
                % [ColdGPT~\cite{cao2023multi}; EmerG~\cite{wang2024warming}
                % ,leaf1,fill=myyellow
                % ]
            %]
            %[Knowledge Graph Relation~\cite{du2022metakg,han2024cold,togashi2021alleviating,pan2024meta},leaf2,fill=myyellow
                % [MetaKG~\cite{du2022metakg}; CRKM~\cite{han2024cold} 
                % ,leaf1,fill=myyellow
                % ]
            %]
        ]
        [Graph Aggregator Improvement,leaf3,edge=black!50, fill=myyellow, minimum height=1.2em,
            [Aggregation Scope Expansion~\cite{hao2021pre,liu2023boosting}; Information Aggregator Augmentation~\cite{hao2023multi,hu2024graph},leaf2,fill=myyellow
                %[Hard Label Guided Contrast 
                %,leaf1,fill=myyellow
                %]
            ]
            %[Homophily Network Relation,leaf2,fill=myyellow
                %[Contrast for Information Fusion 
                %,leaf1,fill=myyellow
                %]
            %]
        ]
    ]
    %%%
    [\textbf{Domain Information} \textcolor{blue}{Section~\ref{sec:domain}},leaf4,edge=black!50, fill=mygreen, minimum height=1.2em
        [Domain Knowledge \\ Transfer,leaf3,edge=black!50, fill=mygreen, minimum height=1.2em,
            [Embedding Mapping,leaf2,fill=mygreen
                [General Mapping~\cite{man2017cross,liu2018transferable,kang2019semi,liu2020attentive,meihan2022fedcdr,bi2020heterogeneous,bi2020dcdir}; Personalized Mapping~\cite{zhu2022personalized,meihan2022fedcdr,vajjala2024vietoris}
                ,leaf1,fill=mygreen
                ]
            ]
            [Heterogeneous Connection,leaf2,fill=mygreen
                [Knowledge Graph~\cite{sun2023remit,zhao2024domain}; Hybrid Connection~\cite{jiang2015social,mirbakhsh2015improving,li2024disco}
                ,leaf1,fill=mygreen
                ]
            ]
            [Learning Process,leaf2,fill=mygreen
                [Learning Techniques~\cite{li2018cross, hu2016learning}; Efficient Tuning~\cite{li2024cdrnp,yi2023contrastive,chen2021user,zhu2021transfer}
                ,leaf1,fill=mygreen 
                ]
            ]
        ]
        [Domain Distribution \\ Alignment,leaf3,edge=black!50, fill=mygreen, minimum height=1.2em,
            [Collaborative Filtering Alignment,leaf2,fill=mygreen
                [Contrastive                 Alignment~\cite{liu2024user,liu2022task,zhao2023cross,xie2024heterogeneous}; Latent-Dimension Alignment~\cite{liu2022task,wang2021low}
                ,leaf1,fill=mygreen
                ]
            ]
            [Auxiliary Feature Alignment,leaf2,fill=mygreen
                [Stein-Path Alignment~\cite{liu2021leveraging,liu2023contrastive}; Contrastive Alignment~\cite{liu2023contrastive,xie2022contrastive} 
                ,leaf1,fill=mygreen
                ]
            ]
        ]
        [Domain-Invariant \\ Representation Learning,leaf3,edge=black!50, fill=mygreen, minimum height=1.2em,
            [Disentangled Representation,leaf2,fill=mygreen
                [Adversarial Learning~\cite{wang2019recsys,su2022cross,wang2024diff,cao2023towards}; Attention Mechanism~\cite{xie2020internal,xie2024heterogeneous,zang2023contrastive}
                ,leaf1,fill=mygreen
                ]
            ]
            [Fusing Representation,leaf2,fill=mygreen
                [Multi-View~\cite{wang2024dual,elkahky2015multi}; Swapping~\cite{cao2022cross,zhao2020catn,wang2020dual};
                Semantics~\cite{wu2020zero,wang2019preliminary,fu2019deeply,he2018general}
                ,leaf1,fill=mygreen
                ]
            ]
        ]
    ]
    %%%
    [\textbf{World Knowledge from LLMs} \textcolor{blue}{Section~\ref{sec:llm}},leaf4,edge=black!50, fill=myred, minimum height=1.2em
        [LLMs as the \\ Recommender System,leaf3,edge=black!50, fill=myred, minimum height=1.2em,
            [Prompting LLMs \\for Recommendation,leaf2,fill=myred
                [Direct Prompting~\cite{sileo2022zero,liu2023chatgpt,hou2024large,sun2024large,sanner2023large,dai2023uncovering,he2023large,wu2024could}; \\Multi-Step Prompting~\cite{wang2023zero,feng2024move,li2024large,di2023retrieval,kieu2024keyword,contal2024ragsys,wu2024coral,liang2024taxonomy,wu2024could,liu2024rec};
                \\Retrieval-Augmented Recommendation~\cite{di2023retrieval,mysore2023large,kieu2024keyword, contal2024ragsys,wu2024coral}
                ,leaf1,fill=myred
                ]
            ]
            [Efficient Tuning LLMs \\ for Recommendation,leaf2,fill=myred
                [Instruction Tuning~\cite{bao2023tallrec,zhang2023collm,zhang2024text,kim2024large,lin2024bridging,geng2022recommendation,chu2023leveraging,xu2024prompting, tan2024idgenrec};\\
                Fine-Tuning~\cite{zhang2024notellm,zhang2024notellm,shen2024exploring,ma2024xrec}
                ,leaf1,fill=myred
                ]
            ]
        ]
        [LLMs as the \\ Knowledge Enhancer,leaf3,edge=black!50, fill=myred, minimum height=1.2em,
            [LLMs for Representation \\ Enhancement,leaf2,fill=myred
                [Modality-Enhanced Representation~\cite{kim2024general,ren2024easyrec,hu2024enhancing}; \\Domain-Enhanced Representation~\cite{tang2023one,gong2023unified,mysore2023large}
                ,leaf1,fill=myred
                ]
            ]
            [LLMs for Relation \\Augmentation,leaf2,fill=myred
                [Behavior Simulation~\cite{huang2024large,wang2024large};\\ External Relation Supplement~\cite{yang2024common,cui2024comprehending} 
                ,leaf1,fill=myred
                ]
            ]
        ]
    ]
]
\end{forest}
\caption{An overview of the taxonomy of this survey for existing cold-start recommendation models.}
\label{fig:overall_structure}
\end{figure}
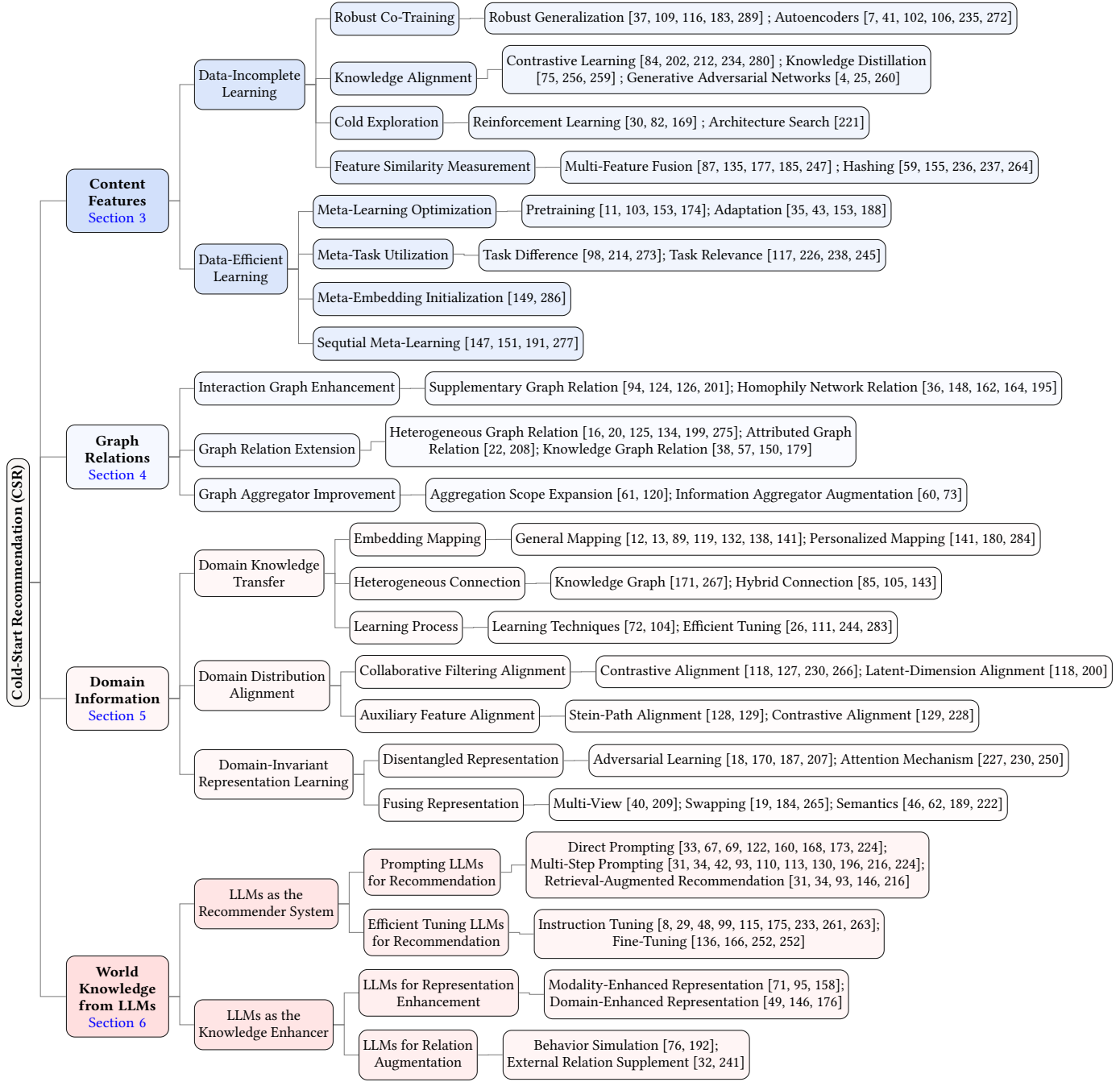

\subsection{Survey Methodology}

To comprehensively cover the papers in the cold-start recommendation. We adopted a semi-systematic survey methodology to identify the relevant papers. Initially, we queried prominent academic databases such as Google Scholar and Web of Science with pre-defined searching keywords such as "cold-start recommendation", "cold-start recommender systems", "strict cold-start", "zero-shot recommendation", and "few-shot recommendation".  
Additionally, we screen specialized conference proceedings, including KDD, WWW, SIGIR, CIKM, WSDM, and RecSys.
The search results were filtered by analyzing titles, abstracts, and experiments to evaluate relevance. Then, the relevant papers were further reviewed thoroughly, and their references were used as seeds for a snowballing approach to identify additional papers. The final collection comprised studies categorized into four core areas based on their contributions, as illustrated in the taxonomy diagram. These areas include content features, graph relations, domain information, and world knowledge from LLMs, as summarized in Figure~\ref{fig:overall_structure}. The majority of these works describe technical approaches or propose novel frameworks, with a smaller subset providing system demonstrations or analytical perspectives on cold-start recommendation methodologies.

\subsection{Contributions}

\begin{itemize}
    \item \textbf{Pioneering Comprehensive Survey:} We present the first thorough review of cold-start recommendation methods, systematically identifying studies from various CSR tasks with different knowledge sources. Our survey meticulously analyzes relevant papers, examining their motivations, data requirements, and technical approaches, providing a consolidated timeline and statistical overview of research publications in leading conferences (e.g., SIGIR, CIKM, KDD) and journals (e.g., TKDE, TOIS), as depicted in Figure~\ref{fig:stat}.
    \item \textbf{Innovative Taxonomy Introduction:} We introduce a novel taxonomy, providing a unique perspective on dealing with the cold-start challenge - utilizing external knowledge sources to address data sparsity and interaction scarcity with new entities. Our taxonomy categorizes knowledge sources distinctly, moving beyond traditional approaches toward a broader scope of addressing cold-start issues.
    \item \textbf{Explicit Definition of Cold-Start Problems:} To the best of our knowledge, we are the first paper offering a clear, comprehensive definition of the cold-start problem across multiple dimensions, encompassing long-tail, user cold-start, item cold-start, user-item cold-start, zero-shot, and few-shot, and strict cold-start problems. This definition provides a structured understanding and unifying framework for diverse research strands within the cold-start landscape.
    \item \textbf{A Foward-Looking Roadmap:} Drawing on our comprehensive survey and innovative taxonomy, we propose a forward-looking roadmap that connects current advancements in cold-start recommendation with future research directions. This roadmap is designed to guide the research community, offering insights and structured pathways for advancing knowledge in this challenging area.
\end{itemize}

\section{PRELIMINARIES}

\subsection{Background}
\subsubsection{Recommender Systems} Recommendation systems (RecSys) are a subclass of information retrieval technologies that seek to predict the preference a user would give to an item or the likelihood of a user's interaction with an item. 
These systems are designed to recommend items to users based on their individual preferences or behaviors, which are mainly inferred from historical user-item interactions. To expand on this, recommender systems play a crucial role in modern e-commerce, social media, and content platforms by helping users navigate through the vast amount of available content and products. They analyze historical behavior data such as purchase history, browsing history, ratings, and reviews to build user profiles and predict what items a user might be interested in~\cite{zhang2024multi,he2020lightgcn,jin2020multi}. 
Specifically, the development of current recommender systems can now be mainly divided into three stages. 
(i) \textbf{Content-based recommendations}. This type of recommender system focuses on the characteristics of items, such as the genre of a movie, the subject of a book, or the style of music. The system recommends items with similar features that a user has liked in the past. 
(ii) \textbf{Collaborative filtering recommendations}. Collaborative filtering (CF) is one of the most commonly used recommendation techniques, which recommends items based on the similarity between users or items, e.g. embedding similarity. User-based collaborative filtering recommends products that other users with similar preferences have liked, while item-based collaborative filtering recommends other items similar to those a user has liked in the past.
(iii) \textbf{Large language model-based recommendations}. In recent years, large language models (LLMs) have received abundant attention for their powerful ability in text-based understanding and generation. Currently, many LLM-based recommender models have been proposed with recommendation-centric prompt tuning and vocabulary extensions for users/items to effectively model the user-item similarity for recommendations.

\subsubsection{Cold-Start Recommendations}
In the above backgrounds of recommender systems, we can find that the core of current recommender models is to mine the user-item similarity with different technical strategies.
However, with the rapid development of the Internet, one major challenge faced by recommender systems is the cold start recommendation (CSR), which involves making accurate recommendations for new users and new items that are continuously added to the Internet every day~\cite{gope2017survey,huang2023aligning,liu2024fine}. The main challenge of cold start recommendation lies in the fact that new users and new items have little or no available information.
In this situation, it is very difficult for the system to model the user-item similarity based on the very sparse information. Therefore, cold-start recommendations have become a long-standing problem for the research community of recommendation systems.
In this survey, we will provide a systematic review of existing CSR methods, starting from the detailed definition of different CSR problems in section~\ref{sec:def} to unfolded classification and discussion of existing CSR models in section~\ref{sec:content} - section~\ref{sec:llm}, with knowledge scopes from content features, graph relations, and domain information, to world knowledge.

\subsection{Problem Definition}\label{sec:def}

\subsubsection{General Problem Definition}

Let $U = \{ u_1, u_2, \dots, u_m \}$ be a set of $m$ users and $V = \{ v_1, v_2, \dots, v_n \}$ be a set of $n$ items. Each user $u \in U$ is associated with a profile $\mathbf{S}_u$, which includes both an interaction history $\mathbf{I}_u$ and contextual features $\mathbf{C}_u$ obtained from external knowledge sources. Similar notations, including the item profile $\mathbf{S}_v$, interactions $\mathbf{I}_v$, and features $\mathbf{C}_v$, hold for each item $v \in V$.
In this setting, the training phase involves a known set of warm-start users $\overline{U}$ and items $\overline{V}$, for which interaction data are fully observed. During tuning and testing, however, we may encounter a new set of cold-start users $\widehat{U}$ and items $\widehat{V}$, which have not been observed during training. By definition, $\overline{U} \cap \widehat{U} = \emptyset$, $\overline{U} \cup \widehat{U} = U$, and similarly $\overline{V} \cap \widehat{V} = \emptyset$, $\overline{V} \cup \widehat{V} = V$. This setup captures the realistic scenario where new users or items emerge after the model is initially trained. Note that for some cases, $\overline{U}$ and $\overline{V}$ could be null due to system-level cold-start in the platform.

\begin{figure}[htbp]
     \centering
     \includegraphics[width=0.915\linewidth, trim=0cm 0cm 0cm 0cm,clip]{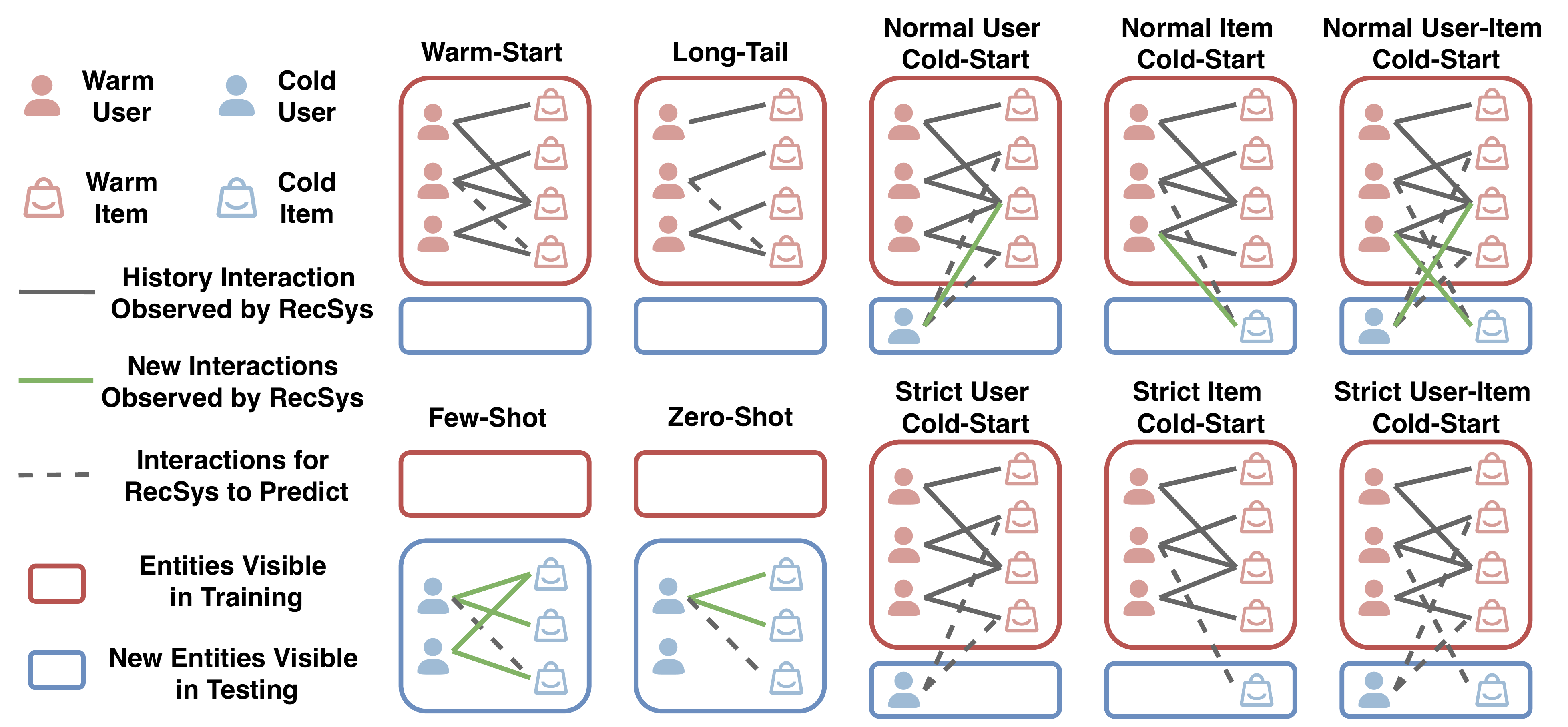}
     \caption{Comparison of warm-start and different cold-start recommendation problems.}
     \label{fig:CSR_problem}
\end{figure}

\subsubsection{Task-Specific Problem Definition}

Building on this general definition, we explicitly define nine specific cold-start recommendation tasks. These tasks differ in terms of the conditions under which users or items are observed by the RecSys, and are grouped into four main categories—long-tail, normal cold-start, strict cold-start, and system cold-start to highlight their unique characteristics. Table~\ref{tab:CSR_problem} and Figure~\ref{fig:CSR_problem} illustrate these categories and the corresponding subtasks, as well as clarify how the training, tuning, and testing sets differ across scenarios.

\begin{table*}[htpb] 
    \centering
    \resizebox{\textwidth}{!}{
    \begin{threeparttable}
    \small
    \caption{Problem definition of different cold-start recommendation tasks.}
    \newcolumntype{H}{>{\setbox0=\hbox\bgroup}c<{\egroup}@{}}
    \begin{tabularx}{\textwidth}{>{\centering}p{1.4cm}|>{\centering}p{1.3cm}|>{\centering}p{.7cm}>{\centering}p{.8cm}>{\centering}p{.8cm}|X}
        \toprule
        {\textbf{Category}} & {\textbf{Task}} & {\textbf{Train}} & \textbf{Tune} & \textbf{Test} & \textbf{Task Specification}\\

        \midrule
        \textbf{Long-Tail} & Long-Tail  & $\overline{U}$, $\overline{V}$ & -- & $\overline{U}$, $\overline{V}$ & Node degrees of users and items in $\overline{U}$ and $\overline{V}$ are less than $k$. \\ 
        \midrule
        \multirow{3}{*}{\makecell{\textbf{Normal} \\ \textbf{Cold-Start}}} & User & $\overline{U}$, $\overline{V}$ & $\widehat{U}$, $\overline{V}$ & $\widehat{U}$, $\overline{V}$  & Each user in  $\widehat{U}$ appear and are observed with less than $k$ interactions.\\ 
                                            & Item  & $\overline{U}$, $\overline{V}$ & $\overline{U}$, $\widehat{V}$ & $\overline{U}$, $\widehat{V}$ & Each item in $\widehat{V}$ are released and observed with less than $k$ interactions.\\ 
                                            & User-Item  & $\overline{U}$, $\overline{V}$ & $\widehat{U}$, $\widehat{V}$ & $\widehat{U}$, $\widehat{V}$ & Users $\widehat{U}$ and Items $\widehat{V}$ appear with less than $k$ interactions each.\\ 
        \midrule
        \multirow{3}{*}{\makecell{\textbf{Strict} \\ \textbf{Cold-Start}}} & User &  $\overline{U}$, $\overline{V}$ & -- & $\widehat{U}$, $\overline{V}$  &  New users $\widehat{U}$ appear after training without any interaction. \\ 
                                            & Item  & $\overline{U}$, $\overline{V}$ & -- & $\overline{U}$, $\widehat{V}$ & New items $\widehat{V}$ are released with no prior interactions.\\ 
                                            & User-Item & $\overline{U}$, $\overline{V}$ & --& $\widehat{U}$, $
                                            \widehat{V}$ & Users $\widehat{U}$ and Items $\widehat{V}$ appear without any user-item interactions each.\\ 
        \midrule
        \multirow{4}{*}{\centering \makecell{\textbf{System} \\ \textbf{Cold-Start}}} & Zero-Shot &  -- & -- & $\widehat{U}$, $\widehat{V}$  & Users $\widehat{U}$ are observed with its short-term interaction histories without collaborative filtering (CF) information from other users and items. \\ 
                                            & Few-Shot  & -- & $\widehat{U}_1$, $\widehat{V}_1$ & $\widehat{U}_2$, $\widehat{V}$ & Users $\widehat{U}_2$ appear with a few interacted items and the learned CF patterns from few-shot users $\widehat{U}_1$ and their interacted $\widehat{V}_1$. \\ 
        \bottomrule
    \end{tabularx}\label{tab:CSR_problem}
    \begin{tablenotes}
        \footnotesize
        \item For splitting $\overline{U}$ and $\widehat{U}$ from $U$, we default to using the number of interactions $k$ in the task specification. Alternatively, some research adopts a time-based approach, categorizing users based on the appearing time (e.g., users' first comment time before year $X$ as existing users and those after as cold-start users), or employs random sampling for the separation. Similarly, these apply to the warm/cold item splitting process.
    \end{tablenotes}
    \end{threeparttable}
    }
\end{table*}

%\clearpage
\section{CONTENT FEATURES}\label{sec:content}
Content features mainly refer to the descriptive information inherent to users or items that characterize their attributes, such as user profiles, user reviews, item names, and descriptions~\cite{acharya2023llm,iaquinta2008introducing,zou2020towards,he2022integrating}. Due to the scarcity or lack of historical interaction records of cold users/items, content features become one of the key information representing cold users/items in cold-start recommendations~\cite{van2013deep,hansen2020content}. Based on the utilization of the content features, we categorize methods into two types: \textbf{Data-Incomplete Learning} (Sec. \ref{sec: inocmplete}), which addresses strict cold-start scenarios without prior interactions, and \textbf{Data-Efficient Learning} (Sec. \ref{sec: efficient}), which optimizes performance in normal cold-start scenarios where limited interaction data is available.

% \begin{figure}[htbp]
%      \centering
%      \includegraphics[width=\linewidth, trim=0cm 0cm 0cm 0cm,clip]{main/images/content-fig.pdf}
%      \caption{Illustrations of the data-incomplete learning with content features, from the perspective of item cold-start.}
%      \label{fig:content_incomplete}
% \end{figure}

\subsection{Data-Incomplete Learning} \label{sec: inocmplete}
Data-incomplete learning is a category of methods that solely utilize content information to learn representations of cold users/items. Given the absence or very few historical interactions for these cold nodes, the available information for modeling their relations is incomplete. Therefore, through data-incomplete learning, which relies only on content information to learn the representations of these strictly cold users/items, the ultimate goal is to unify these representations with those of warm nodes learned from historical interactions for cohesive recommendations.
We categorize the related works of data-incomplete learning into four major classes, \textbf{robust co-training}, \textbf{knowledge alignment}, \textbf{cold exploration}, and \textbf{feature similarity measurement}, based on different learning manners.

\subsubsection{Robust Co-Training}\label{sec:incomplete_robust}
Robust learning is a paradigm in machine learning that aims to make models maintain stability and accuracy even when faced with perturbations, noise, or outliers in the input data~\cite{qayyum2020secure,kang2019robust}.
In cold-start recommendations, robust co-training employs robust strategies to jointly utilize behavior-based warm user/item representations and content-based cold user/item representations for co-training. 
The objective of this training paradigm is to cultivate a model that is not only proficient in leveraging existing behavioral data to refine warm representations but is also adept at warming up cold representations through a process of gradual integration into the training mix.
%The goal of this type of training approach is to enable the model to handle both cold and warm representations by mixing training.
% As the toy example in Figure~\ref{fig:content_incomplete}-(a), robust co-training mainly includes three components: (i) \textbf{Mapper} is a projection function that maps the content information into the embedding representation; (ii) \textbf{Selector} is based on a selection strategy, such as a probabilistic function, to determine whether to train on warm or cold instances at the current stage, which is a key module for the co-training process that combines cold and warm instances. (iii) \textbf{Recommender} is finally adopted to make the recommendation based on the co-trained cold/warm embedding representations. 
Specifically, the models for robust co-training can be divided into two major categories: \textit{\textbf{robust generalization}} and \textit{\textbf{autoencoders}}.

\smallskip\textbf{\textit{Robust generalization}}.
Typically, this type of model will simultaneously optimize both behavior-based representations and content-based representations with robust generalization strategies.
Representatively, DropoutNet~\cite{volkovs2017dropoutnet} randomly selects user-item pairs and sets their preference inputs to zero, forcing the model to rely solely on content information to reconstruct the relevance scores. This approach enables the model to recover the accuracy of the input latent model when preference information is available, while also generalizing in cold situations. 
Heater~\cite{zhu2020recommendation} also employs a similar stochastic training mechanism to randomly input the pre-trained collaborative representations or the intermediate representations derived from contents.
MTPR~\cite{du2020learn} includes both two types of representations: one representation that combines collaborative
embedding and content embedding, and another representation that assumes the item is cold-started, replacing the collaborative embeddings with a zero vector. These two representations are simultaneously used for embedding training. 
Further, there are some models that utilize an adaption-based strategy to transfer knowledge.
For instance, Cold-Transformer~\cite{li2022transform} introduces the context-based embedding adaption for cold/warm users co-training to offset the differences in feature distribution, which transforms the embedding of cold users into a warm state that is more like existing ones to represent corresponding user preferences from historical interaction data.
TDRO~\cite{lin2024temporally} further enhances the generalization capability of the content mapper by integrating time-variant feature shifts within the content feature co-training.

\smallskip\textbf{\textit{Autoencoders}}. The autoencoder (AE) technology employs an encoding-decoding architecture within a unified framework, where the encoder, informed by variational or denoising strategies, and the decoder, responsible for information reconstruction, are jointly trained to represent both cold and warm instances. This synergistic approach effectively captures and reconstructs the nuanced property of the behavior and content data, fostering a robust representation learning process. 
These models generally design new AE architecture to fit both behavior and content information.
Representatively, LLAE~\cite{li2019zero} leverages an AE paradigm that encompasses a low-rank encoder designed to map the user behavior space onto the user content feature space, complemented by a symmetric decoder that reconstructs user behavior from the user content features.
MAIL~\cite{feng2021zero} is also a novel AE, which is
composed of two pivotal autoencoder components: a zero-shot tower to generate behavioral data for cold users with content features and a ranking tower to perform recommendations.
Moreover, CVAR and GoRec~\cite{zhao2022improving,bai2023gorec} introduce a model-agnostic Conditional Variational Autoencoder (CVAE)~\cite{kim2021conditional} framework, which is adept at enhancing the warm-up process for cold-start item ID embeddings.
CFLS~\cite{kong2024collaborative} employs collaborative filtering within the latent variable space by employing a Gaussian process prior to avoiding the difference between behavior embeddings and content embeddings.

\subsubsection{Knowledge Alignment}\label{sec:incomplete_align}
Due to the semantic discrepancies between the warm representations derived from behavioral data and the cold representations obtained from content data~\cite{chen2022generative,huang2023aligning}, a strategic alignment is essential. To bridge this gap, the knowledge alignment introduces strategies to facilitate the convergence of cold representations with the pre-trained warm representations. The core of knowledge alignment is the \textbf{aligner}, which aims to align the cold representations from content features and the warm representations from behavior data with alignment strategies.
These strategies are designed to ensure that the information encapsulated within the cold representations is effectively harmonized with the rich, behavioral-driven insights of the warm representations. By doing so, we aim to enrich the cold representations with the meaningful behavioral information inherent in the warm ones, thereby enhancing the overall semantic coherence and representational fidelity. In terms of technical categorization, existing approaches to knowledge alignment can be divided into three principal categories: \textit{\textbf{contrastive learning}}, \textit{\textbf{knowledge distillation}}, and \textit{\textbf{generative adversarial networks}}. 

\smallskip\textbf{\textit{Contrastive learning}}. It is a famous technique in machine learning that emphasizes the learning of representations by contrasting the similarities and differences between pairs of data points~\cite{zheng2021weakly,chen2020simple}, 
It aims to improve the discriminative ability of models by ensuring that similar instances are represented closely in the representation space, while dissimilar ones are separated~\cite{xie2022contrastive,wu2021self}. In cold-start recommendations, contrastive learning is used to bridge the content-based representations of cold instances with the behavior-based representations of warm instances. For instance, CLCRec~\cite{wei2021contrastive} learns the cold representations from the perspective of information theory, aiming to maximize the interdependence between content information and collaborative signals through contrastive learning. CCFCRec~\cite{zhou2023contrastive} also introduces a contrastive CF framework. It includes a content CF module and a co-occurrence CF module, which work in tandem to produce distinct embeddings. Through joint training and contrastive learning, the co-occurrence signals enhance the content-based embeddings, implicitly correcting the initial imprecision of cold-start item embeddings.

\smallskip\textbf{\textit{Knowledge distillation}}. The knowledge distillation (KD) method is a technique where a smaller, more efficient model learns from a larger, more complex model, transferring the knowledge and insights of the larger model into a more compact form, from the definition in machine learning~\cite{park2019relational,wang2019private,wang2021knowledge,gou2021knowledge}. 
In cold-start recommendations, different from traditional utilization of KD, it is typically adopted to distill the knowledge \textit{from behavior-based warm representations to content-based cold representations}, with the target of ensuring that the representations of the two convey more consistent information.
Representatively, ALDI~\cite{huang2023aligning} views the pre-trained warm recommender model as the teacher and distills knowledge from three perspectives: rating distribution, ranking, and identification, for information alignment between warm representations and cold representations.
Further, DTKD~\cite{zhang2023dual} aims to simultaneously and effectively distill both content and CF knowledge. The DTKD framework employs two specialized teachers - a pre-trained language model and a graph model, each tailored to the distinct characteristics of content and CF data, ensuring comprehensive knowledge distillation. 
Cold \& Warm Net~\cite{zhang2023cold} introduces an expert-driven model that adopts dynamic knowledge distillation as a teacher selector, guiding the experts in refining both cold and warm user representation collaboratively.

\smallskip\textbf{\textit{Generative adversarial networks}}. Generative adversarial networks (GANs) are a class of methods where two neural networks: a generator and a discriminator, contend with each other, enabling the generator to produce increasingly realistic data~\cite{gui2021review,jabbar2021survey,saxena2021generative}. 
In cold-start recommendations, GANs are typically used to make the generated cold representations from the content mapper more similar to the warm representations input into the recommender.
Representatively, GAR~\cite{chen2022generative} employs an adversarial training approach for both the generator (content mapper) and the recommender, ensuring that the generated cold-start item embeddings closely mimic the distribution of warm-start embeddings learned from historical interactions.
GF2~\cite{zhang2022revisiting} also utilizes GANs to enhance the embeddings for cold-start users, with the generator obtained from the GAN that can further be fine-tuned.
GAZRec~\cite{alshehri2022generative} utilizes the GAN to generate virtual representations for cold user/news that are conditioned on the given content data.

\subsubsection{Cold Exploration}\label{sec:incomplete_explore}
In the absence of substantial interaction data to model cold users or items effectively, a natural and intuitive approach is to employ a "trial" methodology, such as reinforcement learning-based strategies~\cite{ji2021reinforcement}. The cold exploration-based approach allows for the exploration of interests among cold users or items, leveraging the feedback signals from the recommender system to swiftly adjust the representations and modeling of these cold entities.

\smallskip\textbf{\textit{Reinforcement learning}}. The reinforcement learning (RL) method is a widely adopted type of exploration method where an agent learns to make decisions by performing actions in an environment to maximize a reward signal~\cite{arulkumaran2017deep,moerland2023model,bei2023reinforcement}. 
For cold-start recommendations, reinforcement learning algorithms are often utilized for interest explorations of cold instances for fast representation cold-starting.
%It's like learning to ride a bike by falling and getting back up, improving over time through trial and error. 
There are some representative methods that model the cold-start process into a specific RL-based process.
Specifically, MetaCRS~\cite{chu2023meta}, through engaging in a series of exploratory dialogues to fast identify cold user preferences for conversational-based recommender systems.
WSCB~\cite{silva2023user} has framed the user cold-start recommendation as a multi-armed bandit problem and introduced an innovative approach to balance exploration and exploitation during initial interactions, which grounds in active learning principles, aims to gather more comprehensive information about cold users.
RL-LTV~\cite{ji2021reinforcement} models the recommendation process as a partially observable and controllable Markov decision process. To transfer the information from historical items to cold-start items, RL-LTV further introduces item inherent features, trending bias terms, and memory states as extra inputs into both the actor and critic.

Some other works further explore the other aspects of the capability of cold-start recommendations. For instance, ColdNAS~\cite{wu2023coldnas} has devised a novel neural architecture search scheme aimed at discovering suitable modulation structures for cold-start users, encompassing both the functional and positional aspects.

\subsubsection{Feature Similarity Measurement}\label{sec:incomplete_feature}
To circumvent the issue of modeling in the face of absent behavioral data, an alternative approach is to shift the focus toward representing and modeling the content-based features of users and items. Specifically, these feature similarity measurement methods learn and evaluate the user/item interests from the perspective of content feature similarity. In this way, the model can avoid the information difference between warm representations (from behavior data) and cold representations (from content data).

\smallskip\textbf{\textit{Multi-feature fusion}}. This method aims to simultaneously utilize multiple features to provide more information for better cold instance measurement due to the data incomplete issue~\cite{yuan2016solving,wang2011collaborative}. The key challenge of this stage is to effectively organize and fuse these features from different sources.
Representatively, CIRec~\cite{ma2023cross} enhances the content representation for cold instances by fusing collaborative, visual, and cross-modal inferred representations.
%incorporates image annotation as privileged information, facilitating the alignment of unified features from the visual domain to the semantic domain during the training phase. Subsequently, it enhances the content representation for cold instances by fusing collaborative, visual, and cross-modal inferred representations.
SMINet~\cite{tao2022sminet} learns the user representations from multiple aspects of input with gate attention to avoid relying solely on behavioral data.
AutoFuse~\cite{jin2023automatic} automatically fuse data from various sources and types, which explicitly categorizes features into distinct groups predicated on their granularity. 
%and learns a multi-tiered representation contingent upon diverse amalgamations of feature groups.

\smallskip\textbf{\textit{Hashing}}. The hashing strategy has been widely adopted in computer vision and multi-media for similar content retrieval to balance the retrieval effectiveness and efficiency~\cite{liu2016deep,wu2019deep,zhu2023multi}. For cold-start recommendations, hashing can be used to map the warm and cold representations into unified binary hash code space for similarity measurement.
For example, NeuHash-CF~\cite{hansen2020content} introduces a content-aware neural hashing method, which generates binary hash codes for both cold and warm users/items, facilitating the estimation of user-item relevance using efficient Hamming distance.
MFDCF~\cite{xu2020multi} proposes a fast cold-start recommendation method with multi-feature discrete CF. Then, it adaptively projects multiple content features into binary and information-rich hash codes for retrieval.

\subsubsection{Others}\label{sec:incomplete_other}
There are also some other methods for data-incomplete learning, which are mainly the early works. Due to the limited quantity, they are mainly based on traditional strategies and statistical methods.
%which are difficult to categorize them into a specific class of models as the above works. 
Representatively, Han et.al~\cite{han2022addressing} combines non-behavioral thematic relevance and behavioral popularity to adjust item rankings, reducing the bias that leads to the cold start ranking of new items.
Deezer~\cite{briand2021semi} utilizes clustering analysis to assign new users to existing user groups, combining user embedding vectors with the centers of user groups to provide recommendations for cold-start users.
DeepMusic~\cite{van2013deep} is a classic cold-start recommendation model that employs mean squared error and prediction error as the objective functions for model training based on rating predictions.

\subsection{Data-Efficient Learning} \label{sec: efficient}

Normal cold-start recommendations are prevalent in many online recommendation systems, prompting another research line to enhance models for efficient learning from limited user-item interactions. Meta-learning, known for its few-shot learning capabilities in fields like computer vision \cite{chen2021meta, jamal2019task}, natural language processing \cite{zhang2022prompt}, and graph mining \cite{zhou2019meta, wang2020graph}, plays a key role here. Gradient-based meta-learning \cite{finn2017model, finn2019online}, which simulates few-shot test scenarios during training and leverages second-order gradients, enables quick adaptation with minimal data. Inspired by these strengths, numerous efforts have focused on applying meta-learning to cold-start problems, categorized into four approaches: \textbf{meta-learning optimization}, \textbf{meta-task utilization}, \textbf{meta-embedding initialization}, and \textbf{sequential meta-learning}.

\subsubsection{Meta-Learning Optimization}\label{sec:eff_learning}
The essence of meta-learning in recommender systems lies in pretraining the model with diverse users’ historical interactions, followed by rapid adaptation to new, cold-start users or items using limited additional interaction data. Thus, refining both the pretraining and adaptation phases is critical for improving cold-start recommendation performance.

\smallskip\textit{\textbf{Pretraining.}} MeLU \cite{lee2019melu} is one of the foundational works in cold-start recommendation, introducing the Model-Agnostic Meta-Learning (MAML) \cite{finn2017model} optimization algorithm to estimate user preferences with only a small number of observed items. During the pretraining phase, local updates are applied to the decision-making layer using the support set, while global updates refine the whole user preference estimator using the query set. The use of second-order gradients on local and global updates enables the model to achieve a robust initial state, allowing it to fine-tune quickly with only a limited number of items to accurately estimate user preferences. Concurrently, building on MAML's strengths, Bharadhwaj \cite{bharadhwaj2019meta} integrated meta-learning strategies as a pretraining enhancement, demonstrating stronger generalization in addressing user cold-start scenarios across various models. Building on these approaches, PNMTA \cite{pang2022pnmta} enlarges the training scope to all the user interactions in pretraining for enriched preference representation learning, while FORM \cite{sun2021form} derives an online regularized meta-leader algorithm and learns from the online gradients to accelerate the meta-training process. 

\smallskip\textit{\textbf{Adaptation.}} 
While much of the focus has been on optimizing global parameter initialization shared for all users, several studies target personalized parameters specifically for cold-start adaptation.
In the early work MAMO \cite{dong2020mamo}, the authors devised two memory matrices to store the feature-specific and task-specific memories for fast preference adaptation. Then Wang et al. \cite{wang2023preference} observed that existing methods risk memorizing query interactions without identifying novel preference patterns and they proposed to separate common preference transfer from novel preference adaptation.
Alternatively, modulators have been proposed for adaptation. For instance, PNMTA \cite{pang2022pnmta} develops extra modulated models tailored for user groups instead of a universal meta-model. It uses a meta-learned task encoder modulator to estimate user identities and a predictor modulator to generate parameters for precise adaptation.
To simplify adaptation pipelines, CMML \cite{feng2021cmml} introduces a fully feed-forward approach. Using a context modulation network, it quickly adapts to limited interactions at both feature and task levels, streamlining the adaptation process.

\subsubsection{Meta-Task Utilization}\label{sec:eff_task}
Beyond optimization, several studies \cite{wen2023modeling, zhao2023task, kim2023meta, yang2022task, lin2021task, yu2021personalized, wu2023m2eu} have highlighted the importance of task similarities and differences in meta-learning. 
Traditional approaches treat each user as an isolated task, training without considering task connections. This limits the model’s ability to recognize individual user contributions and corresponding task relationships. Research in this area focuses on two key aspects: \textit{{task difference}} and \textit{{task relevance}}.

\smallskip\textit{\textbf{Task difference.}} Ignoring task differences can lead to biases, as users with high uncertainty or difficulty may disproportionately affect predictions. 
To address this issue, Wen et al. \cite{wen2023modeling} introduced a weighted distribution of functions (WDoF) framework, using curriculum learning to assign significance to users based on their contribution. 
Zhao et al. \cite{zhao2023task} proposed an adaptive update strategy (TDAS) to distinguish tasks in a macro manner, i.e., covering task difficulty of composition, relevance, and training aspects. Considering the highly distinct user feedback in the real-world cold-start recommendation, Kim et al. \cite{kim2023meta} designed an adaptive weighted loss to capture the imbalanced user rating distribution.

\smallskip\textit{\textbf{Task relevance.}} 
Understanding task relevance is crucial for adapting warm global knowledge to cold-start scenarios, especially when cold users share similar preferences with warm users. Approaches in this domain leverage clustering \cite{yang2022task, lin2021task} or similarity measures \cite{yang2022task,yu2021personalized, wu2023m2eu}.
For instance, Yang et al. \cite{yang2022task} designed an automatic soft task clustering module and a feature-based similarity score to measure task similarity.
In TaNP~\cite{lin2021task}, a clustering distribution is derived through task-adaptive mechanisms to capture task relevance effectively.
Concurrently, Yu et al. \cite{yu2021personalized} also proposed an adaptive meta-learning method focusing on minor users by identifying similar users using a reference tree structure.
Wu et al. \cite{wu2023m2eu} enriched the cold-start user representation by aggregating similar user embeddings through attention similarity scores.

\subsubsection{Meta-Embedding Initialization}\label{sec:eff_embed}
methods focused on adapting models to cold-start scenarios, meta-embedding initialization aims to generate pre-trained embeddings that accelerate the fitting process for cold-start users and items. These methods leverage meta-learning algorithms to produce warmed-up embeddings, enhancing both representation quality and adaptation speed. 
Motivated by the idea of learning better initial embedding, Pan et al. \cite{pan2019warm} proposed to train the meta-embedding generator via a two-phase simulation based on the gradient-based meta-learning \cite{finn2017model}. This approach generates meta-embeddings optimized for strict cold-start conditions, enabling faster adaptation in normal cold-start scenarios. Following that, Zhu et al. \cite{zhu2021learning} proposed a meta-scaling network to transform cold ID embeddings into a warmed ID feature space, accelerating the warm-up process.

\subsubsection{Sequential Meta-Learning}\label{sec:eff_time}
Aligned with the sequential recommendation framework \cite{kang2018self, sun2019bert4rec}, sequential meta-learning incorporates the time order of user interactions to capture dynamic preferences using limited historical behavior sequences.
Wang et al. \cite{wang2021sequential} introduced metric-based meta-learning \cite{vinyals2016matching} paradigm into the sequential recommendation. It focuses on developing a matching network to pair cold-start items with potential users based on limited sequential data.
MetaTL \cite{zheng2021cold} extended gradient-based meta-learning \cite{finn2017model} to sequential recommendations by simulating cold-start scenarios with a pool of few-shot tasks. This setup allows the model to progressively learn user preferences. 
Recognizing that previously active old users may become less engaged over time, Neupane et al. \cite{neupane2022dynamic} defined this group as time-sensitive cold-start users. Their approach dynamically factorizes user preferences into time-evolving representations, combining past and present interactions.
Pan et al. \cite{pan2022multimodal} addressed feature divergence between older and newer interaction sequences. To stabilize and enhance meta-learning, they proposed a Multi-Modal Meta-Learning (MML) framework that integrates diverse side information, such as text and images, to better capture complex user preferences across different types of data.

\section{GRAPH RELATIONS}\label{sec:graph}
%Firstly, here is a brief introduction to graph neural networks, which are widely used for graph relation reasoning in this section:

%Graph Neural Networks. 
In recent years, Graph Neural Networks (GNNs) have captured considerable attention, showcasing cutting-edge performance in a multitude of graph mining tasks, such as node classification~\cite{kipf2017gcn,NIPS2017_graphsage,bei2024cpdg}, link prediction~\cite{zhang2018link,yun2021neo,wu2021hashing}, and graph classification~\cite{xie2022semisupervised,wang2022imbalanced,wei2023neural}. GNNs typically adopt the message-passing paradigm to update each central node embedding via aggregating neighborhood information.
%The widely used message-passing paradigm for GNNs can be described as:
%\begin{equation}
%    \mathbf{e}_{i}^{(k)} = \text{UPDATE}\left(\mathbf{e}_{i}^{(k-1)}, \text{AGGREGATE}(\{\mathbf{e}_{i}^{(k-1)}, \mathbf{e}_{j}^{(k-1)}\}: \forall j \in \gN_{i})\right),
%\end{equation}
%where $\mathbf{e}_{i}^{(k)}$ is the representation of node $i$ in the $k$-th GNN layer, $\gN_{i}$ is the set of neighbors of central node $i$, $\text{AGGREGATE}(\cdot)$ is the message aggregation function from the neighborhood, $\text{UPDATE}(\cdot)$ is the representation updating function for the central node.
As a task within the realm of link prediction, recommender systems have witnessed the emergence of numerous GNN-based recommendation models, which have achieved notable recommendation performance in recent years~\cite{he2020lightgcn,wang2019neural,xu2023graph}. GNN-based recommendation models mainly leverage the powerful message passing of GNNs to model user-item interactions in a graph structure, enabling a better understanding of user preferences and item relevance with high-order information for more effective recommendations~\cite{sharma2024survey,wu2022graph,chen2024macro}. Graph relations provide high-order information, rather than only the content features of the user/item itself. The usage of graph relation knowledge brings information from neighborhoods to a specific user/item. The key challenges in this part lie in \textit{how to provide graph information for cold users/items due to the lack of historical interaction information}. Existing works can be categorized into \textbf{Interaction Graph Enhancement} (Sec. \ref{sec:graph_1}), \textbf{Graph Relation Extension} (Sec. \ref{sec:graph_2}), and \textbf{Graph Aggregator Improvement} (Sec. \ref{sec:graph_3}).
%In the context of a given user $u$ and its associated neighbor item $i$ within a bipartite graph, the formulation of GNN-based models can be succinctly articulated as follows:
%\begin{align}
%&
%\begin{aligned}\label{eq:msg}
%    \mathbf{m}^{(k)}_{u\gets i} = \text{AGGREGATE}(\mathbf{e}^{(k-1)}_{u}, \mathbf{e}^{(k-1)}_{i}),
%\end{aligned}\\
%&
%\begin{aligned}\label{eq:agg}
%    \mathbf{e}^{(k)}_{u} = \text{UPDATE}(\mathbf{m}^{(k)}_{u\gets u}, \sum_{i\in\gN_{u}}\mathbf{m}^{(k)}_{u\gets i}),
%\end{aligned}
%\end{align}
%where $\mathbf{e}^{(k)}_{u}$ denotes the representation of the user node $u$ post $k$ iterations of the GNN message-passing process, capturing insights from its $k$-hop neighborhood. The representation of item $i$ can be obtained in a similar way.

\begin{figure}[tbp]
     \centering
     \includegraphics[width=\linewidth, trim=0cm 0cm 0cm 0cm,clip]{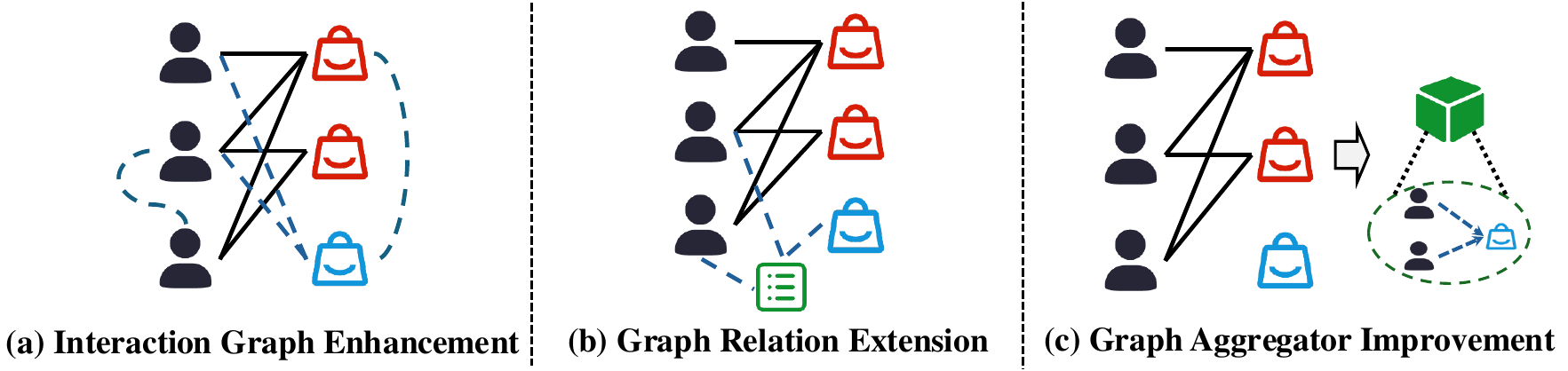}
     \caption{Illustrations of different strategies of graph relation usage.}
     \label{fig:graph}
\end{figure}

\subsection{Interaction Graph Enhancement}\label{sec:graph_1}
Due to the lack of historical interaction behaviors or having very few, providing graph relational information for cold nodes is a significant challenge. Therefore, Interaction Graph Enhancement focuses on increasing the number of interactions on the interaction graph for cold nodes to provide them with more graph information. We categorize the related works of interaction graph enhancement into two major classes: \textbf{supplementary graph relation} and \textbf{homophily network relation}.
The example illustration is shown in Figure~\ref{fig:graph}-(a).

\subsubsection{Supplementary Graph Relation}\label{sec:sup_graph}
This type of model aims to supplement the original user-item interaction graph by including graph relation information for cold instances. 
The key problem for building supplementary graph relations is finding a suitable strategy to generate edges for cold instances and evaluate the quality of the generated edges automatically. Based on the high-quality built edges, the cold instances will have external information aggregated from other nodes. In an ideal way, even warm instances will be positively included information from cold nodes through built interactions.
Representatively, CGRC~\cite{kim2024content} adopts the mask and reconstruction operator on user-item interactions of randomly selected items, enabling the model to infer potential edges for unseen cold start nodes. MI-GCN~\cite{wang2024mutual} enhances the user-item interaction graph with mutual information, where the top similar node pairs under the mutual information evaluation are connected automatically. UCC~\cite{liu2023uncertainty} estimates the uncertainty of each user-item interaction and enhances embedding learning for cold start nodes by adding interactions with low uncertainty.

\subsubsection{Homophily Network Relation}\label{sec:homo_graph}
As the proverb "birds of a feather flock together" suggests, the homophily assumption is a hypothesis often relied upon in graph data mining, indicating that the central node and its neighboring nodes should have similar behaviors or label information~\cite{mcpherson2001birds,ma2022is}. 
To incorporate homophily network relations, algorithms often need to explore explicit/implicit additional associations between users and items, such as social relationships~\cite{li2023survey,sharma2024survey}.
For example, Shams et.al~\cite{shams2021cluster} groups users into clusters based on the homophily similarity of their preferences, thereby accelerating the learning of preferences for new users from warm users. 
GME~\cite{ouyang2021learning} establishes a connection between new items and other relevant existing items through an item graph, and based on this graph, learns how to generate the initial embeddings for new items.
SDCRec~\cite{du2022socially} identifies implicit friend relationships on a user-item-attribute graph by defining palindrome paths, which are based on users having similar evaluations of items.
Recently, Sbandi et.al~\cite{sbandi2024mitigating} simultaneously enhanced cold user–user and item–item link relationships through similarity modeling, resulting in a denser graph for GNN-based recommendations.
%, which are based on users having similar evaluations of items or item attributes.
%PAML~\cite{wang2021preference}; Shams et.al~\cite{shams2021cluster}; GME~\cite{ouyang2021learning}; SDCRec~\cite{du2022socially}

\subsection{Graph Relation Extension}\label{sec:graph_2}
Due to the lack of interaction information, graph relation extension aims to extend the origin interaction graph with more complex relations to pass relevant graph information for cold instances. The methods can be categorized into three classes: \textbf{heterogeneous graph relation}, \textbf{attributed graph relation}, and \textbf{knowledge graph relation}. The example illustration is shown in Figure~\ref{fig:graph}-(b).

\subsubsection{Heterogeneous Graph Relation}\label{sec:hetero_graph} 
Compared to traditional user-item interaction graphs, heterogeneous graphs obtain more complex and information-rich relationships by expanding the types of nodes and edges in the graph. Relationships in heterogeneous graphs are often mined by designing specific heterogeneous graph neural networks~\cite{hu2020heterogeneous,wang2019heterogeneous,zhang2019heterogeneous}.
In the cold start scenario, relying solely on the user-item interaction network cannot meet the demand for relationship mining of cold nodes. Therefore, expanded heterogeneous graphs can often bring more associated information to cold nodes.
The extension is typically based on other available relationships or implicit relationship mining from other information sources. 
Representatively, GIFT~\cite{cao2022gift} establishes a heterogeneous graph that includes physical and semantic links to enhance the message-passing process from preheated videos to cold start videos. 
Further, HGNR~\cite{liu2020heterogeneous} constructs a heterogeneous graph that is composed of user-item interactions, social links, and semantic links predicted from social networks and textual reviews. 
%Physical links represent explicit relationships, while semantic links measure the closeness of the multimodal representations of two videos.
MvDGAE~\cite{zheng2021multi} enhances the connections between users and items in different aspects through multi-view extraction.
PGD~\cite{wang2021privileged} and IHGNN~\cite{cai2023user} incorporate attribute information of users and items into the user-item graph to construct a heterogeneous graph (user-item-attribute graph), enabling cold nodes to have more available information for aggregation within this graph. %IHGNN~\cite{cai2023user} also constructs a user-item-attribute graph for recommendation to avoid the link sparse issue of cold instances.
%HGNR~\cite{liu2020heterogeneous}; MetaHIN~\cite{lu2020meta}; PGD~\cite{wang2021privileged}; MvDGAE~\cite{zheng2021multi}; GIFT~\cite{cao2022gift}; IHGNN~\cite{cai2023user}

\subsubsection{Attributed Graph Relation}\label{sec:attri_graph}
Attributes typically reveal the inherent information of an instance, and similar attributes can represent that the two have similar characteristics, such as items with similar descriptive information may belong to the same category or users with similar profiles may be part of the same interest community~\cite{he2018nais,mcfee2012learning}. 
In cold start recommendations, attribute graphs are also often used for message passing to avoid the issue of having no available interaction information. Specifically, ColdGPT~\cite{cao2023multi} leverages LLMs to extract fine-grained attributes from item content and connect them to item nodes to form an item-attribute graph structure for cold-start representation learning of items. EmerG~\cite{wang2024warming} builds an item-specific feature graph with a GNN message passing on it to conduct CTR prediction with cold items.

\subsubsection{Knowledge Graph Relation}\label{sec:kg}
A knowledge graph (KG) is a structured semantic knowledge base that stores relationships between entities in the form of a graph, with nodes representing entities and edges representing various semantic relationships between entities~\cite{guo2020survey,ji2021survey}. 
%This type of graph provides rich information, supports complex queries and reasoning, and is widely used in search engines, recommendation systems, and intelligent question-answering fields~\cite{guo2020survey,ji2021survey}. 
The auxiliary information in knowledge graphs can be utilized to enhance cold instance learning.
Representatively, KGPL~\cite{togashi2021alleviating} leverages unobserved user-item pairs as weak positive or negative instances, assigning pseudo-labels to these unobserved samples.
%through model predictions rather than merely treating them as negative instances. 
To enhance the accurate labeling of cold-start users through pseudo-labeling, KGPL conducts sampling based on the structure of the knowledge graph, selecting items that may potentially interact positively with users.
MetaKG~\cite{du2022metakg} includes two meta-learners: a collaborative sensing meta-learner and a knowledge sensing meta-learner. These two learners respectively capture user preferences and knowledge of KG entities to combine more information for adapting to cold-start recommendations.
CRKM~\cite{han2024cold} utilizes knowledge graphs and popularity information to sample negative labels from cold items that have not interacted with users, thereby alleviating the sparsity of cold-start training data.

\subsection{Graph Aggregator Improvement}\label{sec:graph_3}
The two aforementioned subsections (interaction graph enhancement and graph relation extension) primarily address the issue of structural information scarcity for cold nodes by enhancing the graph structure. Another approach is to design an augmented model that extracts more usable information from limited structural data for cold-start recommendations, which we call graph aggregator improvement with a model-centric perspective in this survey.
Related works can be categorized into two classes: can be mainly categorized into two main approaches: \textbf{expanding the aggregation scope} and \textbf{augmenting the information aggregator}.
The example illustration is shown in Figure~\ref{fig:graph}-(c).

\subsubsection{Aggregation Scope Expansion} The first approach extends the model's scope beyond local neighborhoods, encouraging attention to global or long-range contexts. In this way, cold instances can perceive long-distance correlated nodes to alleviate the sparsity in the direct neighborhood. For example, MeGNN~\cite{liu2023boosting} employs global neighborhood transformation learning to achieve consistent latent interactions for all new users and item nodes and adopts local neighborhood transformation learning to forecast specific latent interactions tailored to each node. MPT~\cite{hao2023multi} integrates a Transformer encoder into the GNN encoder framework to capture long-range dependencies between users and items, thereby providing cold nodes with access to a richer set of usable neighborhood information.

\subsubsection{Information Aggregator Augmentation} Meanwhile, the second approach refines the aggregator's functionality, enabling it to capture more critical information for cold nodes within the limited interaction data of cold instances. Representatively, to mitigate the impact of cold-start neighbors, Hao et.al~\cite{hao2021pre} introduces a meta-aggregator based on self-attention to enhance the aggregation capabilities at each graph convolution step.
A-GAR~\cite{hu2024graph} introduces an adaptive neighbor aggregation strategy, comprehensively exploring higher-order features of users/items. Based on this, a graph attention network is employed to integrate the augmented preference information from neighbors, enhancing aggregators' ability to model data sparsity in cold-start scenarios.

\section{DOMAIN INFORMATION}\label{sec:domain}
\begin{figure}[htbp]
     \centering
     \includegraphics[width=\linewidth, trim=0cm 0cm 0cm 0cm,clip]{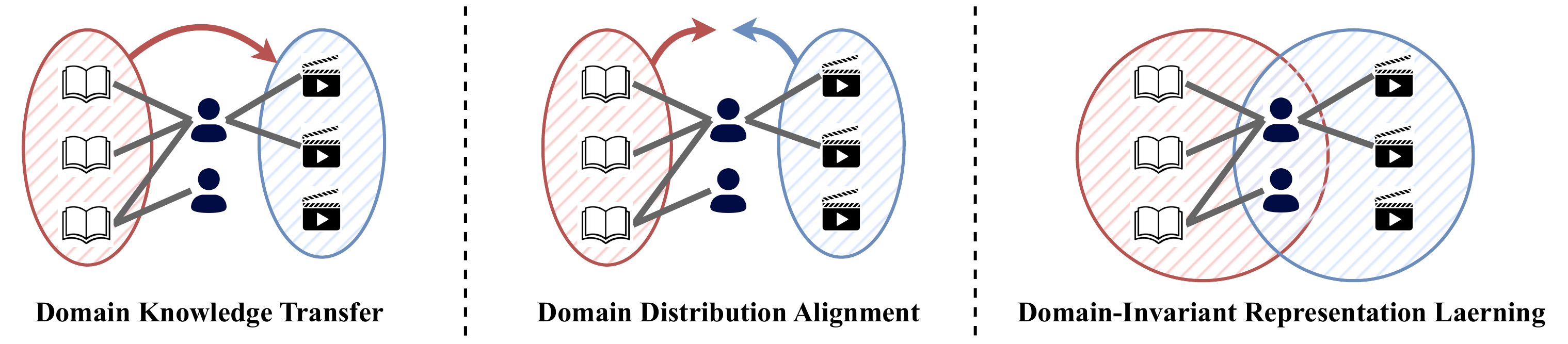}
     \caption{Illustrations of different categories of methods for utilizing the cross-domain knowledge.}
     \label{fig:domain}
\end{figure}

In real-world online applications, only a few platforms experience significant user engagement, while many others struggle with persistent long-tail and user cold-start issues. Therefore, transfer learning~\cite{weiss2016survey,zhuang2020comprehensive} across different domains offers a promising solution by leveraging knowledge from source domains with abundant data to enhance recommendation performance in target domains with limited information. Unlike traditional cold-start recommendation systems, cross-domain recommendation methods are inherently more complex. They must consider knowledge from at least two distinct systems, which often differ significantly. These methods generally require overlapping users in cross-domain settings and strategies to effectively utilize those users to share domain knowledge. According to the high-level methodologies of utilizing the domain knowledge, we divide existing work into three classes: \textbf{Domain Knowledge Transfer} (Sec. \ref{sec:domain_tranfer}), \textbf{Domain Distribution Alignment} (Sec. \ref{sec:domain_align}), and \textbf{Domain-Invariant Representation Learning} (Sec. \ref{sec:domain_invariant}) as illustrated in Figure~\ref{fig:domain}.

\subsection{Domain Knowledge Transfer}\label{sec:domain_tranfer}
Domain transfer methods provide a straightforward approach to tackling cold-start problems in cross-domain scenarios. These methods typically rely on \textit{\textbf{embedding mapping}}, \textit{\textbf{graph connections}}, or \textit{\textbf{learning processes}} to facilitate the seamless transfer of knowledge from a warm source domain to a cold target domain.

\subsubsection{Embedding Mapping}
One simple way is generalizing to the cold-start domain via various embedding mapping and feature transfer techniques. These methods typically focus on aligning the embedding spaces of the two domains through non-linear transformations, ensuring a smooth and effective knowledge transfer process.

\smallskip\textit{\textbf{General Mapping.}} Much of the work \cite{man2017cross,liu2018transferable,kang2019semi,liu2020attentive,meihan2022fedcdr,bi2020heterogeneous,bi2020dcdir} simply adopts a single multi-layer perceptron (MLP) for general mapping of two domain representation space.  
For collaborative filtering-based mapping, the assumption is that overlapping users have rich interaction data. Studies like \cite{man2017cross, liu2018transferable, kang2019semi} leverage MLPs for flexible, non-linear transformations of the overlapping users' representation spaces across domains.
For side-information-based mapping, MAFT \cite{liu2020attentive} utilizes MLPs combined with attention mechanisms to map auxiliary feature spaces effectively. In contrast, HCDIR \cite{bi2020heterogeneous} and DCDIR \cite{bi2020dcdir} enhance the target domain's semantic and informational modeling by constructing heterogeneous information networks.

\smallskip\textit{\textbf{Personalized Mapping.}} Instead of employing a universal mapping function, recent methods \cite{zhu2022personalized, meihan2022fedcdr, vajjala2024vietoris} focus on personalized mapping tailored for each user.
For example, PTUPCDR \cite{zhu2022personalized} introduces a meta-network that generates personalized parameters for bridging functions by using user-specific embeddings learned from the source domain.
In comparison, VRCDR \cite{vajjala2024vietoris} incorporates Vietoris-Rips complexes, using characteristic vectors derived from users' interaction patterns as inputs for the mapping function. This method models user preferences and geometric relationships between interacted items, translating user embeddings from source to target domains through personalized vectors.

\subsubsection{Heterogeneous Connections}
Beyond constructing a single mapping function to connect domains, heterogeneous approaches \cite{sun2023remit, zhao2024domain, jiang2015social, mirbakhsh2015improving} establish richer connections to facilitate domain knowledge transfer. These methods leverage auxiliary graph network structures to explicitly model and transfer knowledge across domains. 

\smallskip\textit{\textbf{Knowledge Graph.}} Based on the rich information in the knowledge graph, \cite{sun2023remit} uses meta-path-based aggregation and multiple personalized bridges for transforming interest embeddings. On the other hand, \cite{zhao2024domain} focuses on cross-domain knowledge graphs by capitalizing on natural relationships between items across domains, such as books and their movie adaptations.

\smallskip\textit{\textbf{Hybrid Connections.}} Another option is to construct hybrid connections over users or items in different domains. Jiang et al. \cite{jiang2015social} proposed to model a social network as a star-structured hybrid graph, where a central social domain connects with multiple item domains (e.g., web posts, and videos). CBMF~\cite{mirbakhsh2015improving} utilized clustering-based relations to capture shared interests between user/item groups across domains while DisCo~\cite{li2024disco} introduces multi-channel graph encoders to obtain diverse user intents.

\subsubsection{Learning Process}
In the meanwhile, researchers have resorted to different training and tuning techniques to implicitly pass information from warm/source domains to cold/target domains.

\smallskip\textit{\textbf{Training Techniques.}}
Early cross-domain works \cite{hu2016learning, li2018cross} proposed specialized learning techniques to facilitate joint training across domains. The WITF model \cite{hu2016learning} trains on multi-domain feedback to learn informative priors about users and items to regularize the user preferences inference process. \cite{li2018cross} developed a rating prediction model based on Partial Least Squares Regression that can transfer users' rating preferences effectively. 

\smallskip\textit{\textbf{Efficient Tuning.}} 
Recently, {efficient tuning} strategy \cite{li2024cdrnp,yi2023contrastive,chen2021user,zhu2021transfer} has gained popularity in the cross-domain cold-start recommendation. As the overlapping users might be limited across domains, \cite{zhu2021transfer,li2024cdrnp} leverage meta-learning-based tuning approaches \cite{finn2017model} to feature a two-stage process — pre-training models on source and target domains followed by meta-learning to tune a task-specific meta-network that generalizes knowledge for cold-start.
Following them, \cite{li2024cdrnp} applies neural processes (NP) within the meta-learning tuning paradigm to model user-specific preferences and capture preference correlations among overlapping and cold-start users by representing preferences as predictive probability distributions.
Other than meta-learning, Chen et al. \cite{chen2021user} proposed a User-specific Adaptive Fine-tuning (UAF) to enhance personalization, and Yi et al. \cite{yi2023contrastive} utilized contrastive learning and personalized prompts to transfer knowledge from a source domain to a target domain.

\subsection{Domain Distribution Alignment}\label{sec:domain_align}
Domain alignment in cold-start RecSys focuses on reducing distributional differences between source and target domains to enable effective knowledge sharing. By aligning shared features, user behaviors, or auxiliary information across domains, these methods address challenges like data sparsity and cold-start scenarios.

\subsubsection{Collaborative Filtering Alignment}
The objective of collaborative filtering (CF) alignment is to leverage shared interaction patterns and user behaviors to facilitate better knowledge transfer among domains. To bridge distribution gaps, various methods \cite{liu2024user,xie2024heterogeneous,zhao2023cross,liu2022task,wang2021low} have been developed to ensure that the CF models can generalize well on the cold users in target domains in the normal user cold-start or long-tail settings. 

\smallskip\textit{\textbf{Contrastive Alignment.}} Contrastive learning has been widely used to address distributional differences and fulfill the domain alignment goal.
%www24
For example, \cite{liu2024user} integrates a rating prediction module and a distribution alignment module, where the latter aligns both overlapped and non-overlapped users across domains using unbalanced distribution optimal transport with contrastive loss. 
%kdd22
Such sample-wise contrastive alignment is also implemented in DAUC~\cite{liu2022task} to address the domain distribution gap.
%www23 contrast
Focusing on user interest alignment, \cite{zhao2023cross} constructs a unified cross-domain heterogeneous graph to capture high-order user-item relationships across domains. This model uses contrastive learning and gradient alignment to align user-user and user-item interest spaces effectively.
%kbs24
Building on this, HGCCDR \cite{xie2024heterogeneous} constructs a heterogeneous graph enriched with user ratings, reviews, and item categories. Through graph augmentation and contrastive learning, HGCCDR strengthens both intra- and inter-domain connections, optimizing user embeddings for cross-domain scenarios.

\smallskip\textit{\textbf{Latent-Dimension Alignment.}} This line of studies \cite{liu2022task,wang2021low} leverages the encoder-decoder architecture to align domains in the latent representation space.
%kdd22
DAUC~\cite{liu2022task} addresses the domain distribution gap between mobile app usage and article reading by using contrastive alignment and adversarial alignment loss on latent dimensions in the encoder-decoder structures.
%cikm21
Similarly, LACDR \cite{wang2021low} maps user embeddings to a low-dimensional space and aligns overlapping user representations to extract domain-invariant features.

\subsubsection{Auxiliary Feature Alignment}
Aligning latent embedding distributions directly between source and target domains is inherently challenging. Enforcing the alignment of two domains on the common auxiliary features space makes the process more approachable. 

\smallskip\textit{\textbf{Stein-Path Alignment.}} Representatively, inspired by the idea of particle-based inference in stein variational gradient descent~\cite{han2018stein}, DisAlign \cite{liu2021leveraging} adapts target domain samples to align with source domain distributions by iteratively moving embeddings along probabilistic Stein paths. For example, a book's auxiliary embedding can align with its corresponding movie adaptation in a target domain, forming a semantic bridge. Expanding on this concept, CPKSPA \cite{liu2023contrastive} introduces proxy Stein path alignment to further reduce domain gaps by moving cold item embeddings with their source domain counterparts.

\smallskip\textit{\textbf{Contrastive Alignment.}}
Paired with contrastive augmentation and alignment, CPKSPA~\cite{liu2023contrastive} further enhances robustness through contrastive learning of cold item embeddings.
Similarly, CCDR \cite{xie2022contrastive} uses contrastive learning to address domain alignment through both intra- and inter-domain techniques. Intra-domain contrastive learning creates augmented sub-graphs to combat data sparsity, while inter-domain learning aligns users, taxonomies, and neighbors to improve cross-domain knowledge transfer. Together, these methods create a comprehensive framework for aligning auxiliary feature spaces to enhance the cold-stat recommendation.

\subsection{Domain-Invariant Representation Learning}\label{sec:domain_invariant}
Instead of focusing on reducing the distributional differences as in domain alignment, domain-invariant representation learning assumes there is a shared feature space that are universally transferable across domains, capturing common user preferences or item characteristics.

\subsubsection{Disentangled Representation}
The common optimization goal of approaches in this category is to separate domain-invariant (shared) and domain-specific features. During training, the disentangling process ensures that shared representations capture universal user preferences or item characteristics that are transferable across domains, while domain-specific representations retain unique traits relevant to individual domains.

\smallskip\textit{\textbf{Adversarial Learning.}} This technique has been widely adopted to disentangle domain-invariant and domain-specific features.
% cikm22 tnnls20
Following the adversarial generative network paradigm \cite{goodfellow2014generative}, RecSys-DAN~\cite{wang2019recsys} and AA~\cite{su2022cross} train discriminators to distinguish between source and target domain representations, while generators produce domain-invariant features that confuse the discriminator.
Diff-MSR \cite{wang2024diff} enhances this approach by using diffusion models to generate embeddings for each domain, isolating domain-shared and domain-specific characteristics via a classifier. UniCDR \cite{cao2023towards} further incorporates adversarial learning with contrastive objectives and masking mechanisms to perturb interactions and domain settings, generating diverse and robust representations.

\smallskip\textit{\textbf{Attention Mechanism.}} It offers another pathway for disentanglement by focusing on feature-field relationships.
\cite{xie2020internal} uses a multi-channel attention mechanism with contextual and internal attention layers to extract domain-specific and domain-invariant features. Building on this, \cite{xie2024heterogeneous} employs a multi-layer attention mechanism over user-item heterogeneous graphs, aligning domain-shared features using cross-view contrastive learning. In \cite{zang2023contrastive}, attention is combined with contrastive learning to separate domain-invariant knowledge from domain-specific noise, using one attention module to aggregate shared knowledge and another to isolate domain-specific features.

\subsubsection{Fusing Representation}
The key idea is to fuse features between domains via multi-view learning and swapping learning, enabling the models to generalize domain-invariant user behaviors across different contexts.

\smallskip\textit{\textbf{Multi-View Learning.}}
In short, multi-view learning leverages complementary perspectives to create shared representations for cross-domain recommendations. For instance, \cite{wang2024dual} explores dual-view learning by combining content semantics (similarities between items based on descriptive features) and structural connectivity ( high-order relationships in knowledge graphs) to achieve domain-invariant representations. Similarly, \cite{elkahky2015multi} utilizes multi-view learning to integrate features, such as user behavior data (e.g., searches, clicks) and item characteristics (e.g., apps, movies, news), into a shared latent space to enhance knowledge transfer.
% % tkdd24 www15

\smallskip\textit{\textbf{Swapping Learning.}}
The swapping strategy refines domain-invariant features by exchanging domain-specific information. CDRIB \cite{cao2022cross} employs variational information bottleneck principles to swap and balance domain-shared and domain-specific features, thus filtering out noisy, irrelevant features. CATN \cite{zhao2020catn} swaps aspect-level preferences, such as plot or genre, extracted from reviews across domains (e.g., books to movies), enabling effective domain-invariant feature extraction. Differently, Dual Autoencoder Network (DAN) \cite{wang2020dual} employs a swap reconstruction strategy. Specifically, the model uses dual encoder-decoder networks where user representations are swapped between the source and target domains for reconstruction. In this swap process, representations from one domain are reconstructed in another, ensuring that cross-domain information is leveraged effectively.

\smallskip\textit{\textbf{Semantic Learning}}
Approaches in semantic learning aim to bridge domains by fusing different auxiliary information from users and items and assuming the semantic features space as domain-invariant. Earlier attempts like \cite{he2018general} and \cite{wang2019preliminary} map auxiliary information (e.g., user reviews, browsing histories, item descriptions) into a shared semantic space, enabling knowledge transfer. Following them, RCDFM \cite{fu2019deeply} uses Stacked Denoising Autoencoders (SDAEs) to fuse semantic representations from user reviews and item contents with rating matrices, creating richer latent factors. Furthermore, \cite{wu2020zero} introduces zero-shot heterogeneous transfer learning to align semantic spaces between a recommender system and a retrieval system, leveraging item co-consumption correlations to generate domain-invariant embeddings.

\section{WORLD KNOWLEDGE FROM LARGE LANGUAGE MODELS}\label{sec:llm}
\label{llm}
Large language models (LLMs) are generative artificial intelligence systems trained using deep learning techniques to understand the general world knowledge by learning vast amounts of textual corpus data. These models can generate text, answer questions, perform translations, and even engage in complex conversations~\cite{zhao2023survey,zhu2023large}. Due to the tremendous success achieved in recent years, an increasing number of fields have begun to leverage the capabilities of large language models for various tasks, such as multimodal learning~\cite{wu2023multimodal}, graph learning~\cite{ren2024survey}, and recommender systems~\cite{wu2024survey}, achieving commendable results.
Due to the powerful textual feature processing capabilities of LLMs, cold start, especially the zero-shot and few-shot scenarios, has become an important application in the recommendation domain for LLMs. According to the role that LLMs play, we categorize existing works into two main aspects: \textbf{LLM as the Recommender System} (Sec. \ref{sec:llm_as_rec}) and \textbf{LLM as the Knowledge Enhancer} (Sec. \ref{sec:llm_as_enhancer}).

\begin{figure}[htbp]
     \centering
     \includegraphics[width=\linewidth, trim=0cm 0cm 0cm 0cm,clip]{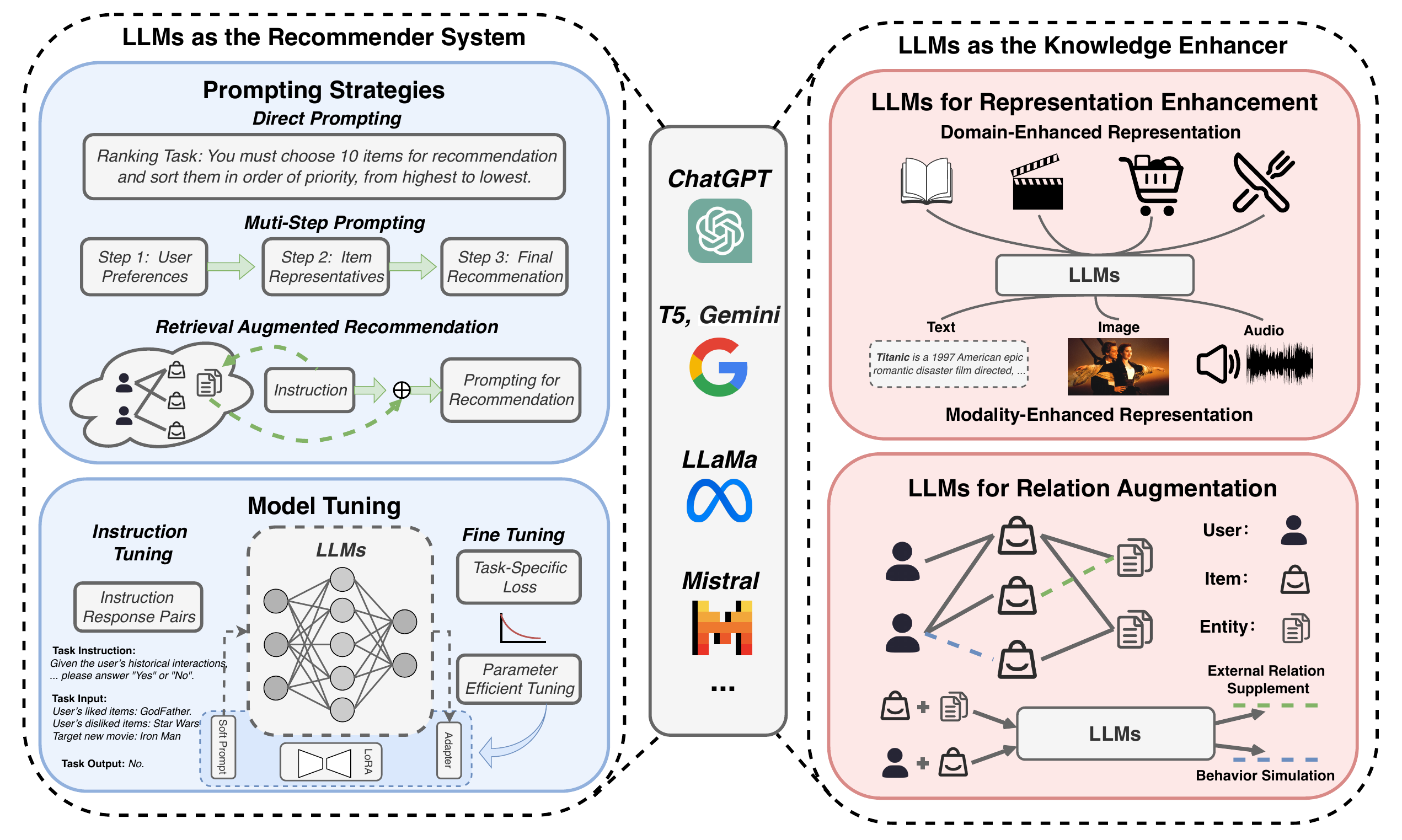}
     \caption{Illustrations of different categories of methods for utilizing the world knowledge from LLMs.}
     \label{fig:llms}
\end{figure}

\subsection{LLM as the Recommender System}\label{sec:llm_as_rec}
Given the remarkable advancements of large language models (LLMs) in various zero-shot natural language processing tasks, it is natural to explore their potential for cold-start recommendations, especially in zero-shot scenarios. In these cases, LLMs can provide personalized suggestions without the need for domain-specific fine-tuning or extensive training on historical user interactions. This line of research seeks to address the limitations of traditional recommender systems that heavily rely on historical interactions by leveraging the contextual understanding and generative capabilities of LLMs in tasks such as next-item prediction, conversational recommendation, and explainable recommendation.

\subsubsection{Prompting Strategy}
This series of work designed different prompting strategies to guide LLMs in making accurate (system) cold-start recommendations by framing recommendation tasks as natural language processing problems. This involves techniques like designing direct task-specific prompts including demonstrating in-context examples for the recommendation, integrating multi-step prompting for external information (such as taxonomy dictionaries or image summaries), and retrieval-augmented information to adapt LLMs to rank or generate recommendations based on the interactions and content features of users and items. 

\smallskip \textbf{\textit{Direct Prompting.}}
%ecir22 
\cite{sileo2022zero} first presented a simple direct prompting method where cold-start recommendations are generated using pretrained language models (e.g., GPT-2) without any need for task-specific retraining. 
% cikm23
With the update of ChatGPT, \cite{liu2023chatgpt} explores the use of GPT-3.5 in zero-shot/few-shot recommendation tasks such as rating prediction, sequential recommendation, direct recommendation, explanation generation, and review summarization. 
% ecir24 Arxiv24 RecSys23_2
Concurrently, some similar work \cite{hou2024large,sun2024large,sanner2023large,dai2023uncovering} employ task-specific prompting to directly conduct recommendation tasks (e.g., formulating specific prompts for tasks like rating prediction and item recommendation). 
% ecir24 Arxiv24 RecSys23_2
Based on the task instruction, they further utilize few-shot in-context learning to incorporate user-item interaction to further guide recommendations).
%cikm23_2, Arxiv24
\cite{he2023large,sun2024large} performed conversational recommendations in the cold-start settings and employed a task-specific prompting strategy to guide LLMs to generate responses based on the conversation. 
%www24
Notably, PromptRec~\cite{wu2024could} proposed the extreme system cold-start scenarios where no historical user-item interactions are available (e.g., new businesses). They cope with the problem by reframing the recommendation task into a sentiment analysis task using user and item profiles expressed in natural language.

\smallskip \textbf{\textit{Muti-Step Prompting.}}
Though direct prompting can be used for zero-shot/few-shot recommendation tasks, they demonstrate less competitive results than traditional methods fully trained on user-item interactions. 
To improve the recommendation capabilities of LLMs, researchers attempt to restructure and transform originally complex and ambiguous cold-start recommendation tasks into more manageable muti-step tasks with modality-rich information.
% arixiv23
Representatively, \cite{wang2023zero} involves a 3-step prompting approach using GPT-3 to (1) capture user preferences, (2) select representative movies, and (3) recommend items based on these inputs. This approach innovatively designs a multi-step pipeline to guide GPT-3 through subtasks to improve recommendation accuracy.
% cai24
Similarly, \cite{feng2024move} applies a 3-step prompting where the LLM is guided through preprocessing background data, weighing user behavior factors, and generating POI recommendations with explanations.
Some work attempts to adopt extra steps to gather more information before prompting the LLMs.
For instance,
% sigir24
\cite{li2024large} first transform the raw check-in records of users into question-answering tasks via key-value similarity and then specifically design prompts using the current and past trajectory along with the task instruction.
% recsys23_3  % arxiv24_3 % kieu2024keyword, contal2024ragsys,wu2024coral RAG
Furthermore, the taxonomy dictionary is integrated into the overall multi-step recommendation framework to gather relevant information including item genres and item themes in \cite{liang2024taxonomy}.
% emnlp23 arxiv24_2
Beyond the extra steps for prompting or textual information collection, visual data has been utilized in \cite{wu2024could,liu2024rec} via few-shot in-context learning. Specifically, \cite{wu2024could} introduces human-like explanations of visual features for LLM-based recommendations while \cite{liu2024rec} introduces visual-summary thought (VST), a reasoning strategy for summarizing the textual descriptions of images in multimodal recommendation tasks.

%di2023retrieval,mysore2023large,kieu2024keyword, contal2024ragsys,wu2024coral
\smallskip \textbf{\textit{Retrieval-Augmented Recommendation.}}
In the field of NLP, retrieval-augmented generation (RAG) is an effective solution that enriches LLMs with a comprehensive array of background knowledge and detailed contextual insights to significantly enhance their capacity for content generation tasks~\cite{zhao2024retrieval,gao2023retrieval}. 
In recommendation systems, the retrieval-augmented modules~\cite{di2023retrieval,mysore2023large,kieu2024keyword, contal2024ragsys,wu2024coral} are primarily utilized to retrieve related exact item entities or essential information for encoding users/items, thereby enhancing the modeling accuracy of LLM-based recommender models. By leveraging the additional and fine-grained information obtained through retrieval, the modeling of cold and zero-shot instances, which initially lacked sufficient data, can be significantly strengthened for recommendations.
Most retrieval-based recommendation methods first conduct the retrieval process for collecting relevant user/item information such as item characteristics~\cite{di2023retrieval}, potential item candidates \cite{kieu2024keyword}, and users/items with corresponding interaction history \cite{contal2024ragsys,wu2024coral} and then prompting the LLM-based recommender system with such information to facilitate recommendation.
In contrast, Mint~\cite{mysore2023large} initially utilizes large language models (LLMs) to generate synthetic narrative queries through few-shot prompting and trains retrieval models on these synthetic queries and user-item interaction data. The core of the Mint method lies in repurposing the rich dataset of user-item interactions via LLMs for training the retrieval model to enhance the performance of the recommendation system.

\subsubsection{Model Tuning}
The motivation for tuning LLMs arises from the need to bridge the gap between the preference-capturing process of RecSys and the rich semantic understanding of LLMs in (system) cold-start recommendation. Though effective in zero-shot settings, direct-prompting-based LLMs struggle to handle and transfer rich collaborative filtering information from warm users/items to cold ones. In addition, they only achieve competitive cold-start performance at the cost of sacrificing the warm-start recommendations. To address these limitations, recent works \cite{bao2023tallrec,zhang2023collm,zhang2024text,kim2024large,lin2024bridging,geng2022recommendation,chu2023leveraging,xu2024prompting, tan2024idgenrec, zhang2024notellm,zhang2024notellm,shen2024exploring,ma2024xrec} focus on learning interaction patterns through the tuning process. This approach enables LLMs to incorporate their pre-trained world knowledge and collaborative filtering interactions into the recommendation decision process, facilitating both warm-start and cold-start recommendations more effectively.

%bao2023tallrec,zhang2023collm,zhang2024text,kim2024large,lin2024bridging,geng2022recommendation,chu2023leveraging,xu2024prompting, tan2024idgenrec, tsai2024leveraging 
% tsai2024leveraging
\smallskip\textit{\textbf{Instruction Tuning.}}
The general framework of instruction tuning for cold-start LLM RecSys is to convert recommendation tasks into text-to-text generation processes, where the model learns to interpret user histories and item metadata in natural language via the next token prediction training. The constructed instruction-based dataset enables LLM RecSys to gain collaborative filtering knowledge in recommendation and seamless adapt to new items and domains without the need for additional time-consuming pre-training.
% recsys23
Among this category, TALLRec~\cite{bao2023tallrec} is one of the most pioneering works focusing on instruction tuning for RecSys. In particular, the original recommendation data is structured as natural language recommendation instructions to guide the LLMs to answer binary classification outputs ("yes" or "no").
% recsys22
Another fundamental work in instruction tuning for LLM-based recommendation is P5 (Pretrain, Personalized Prompt, and Predict Paradigm)~\cite{geng2022recommendation}. It proposed a unified framework for various recommendation tasks, where all types of recommendation data (user-item interactions, metadata, and reviews) are converted into natural language prompts. This method allows P5 to handle different recommendation tasks in a shared text-to-text encoder-decoder framework without relying on task-specific training.
% arxiv23 KDD24 acl24 
Building on these two works, a series of studies have been proposed to further enhance the utilization of collaborative filtering signals among warm users and items during the LLM tuning stage.
For instance, CoLLM~\cite{zhang2023collm} and A-LLMRec~\cite{kim2024large} leverage a mapping/alignment module to inject collaborative information from an external collaborative model. BinLLM~\cite{zhang2024text} aims to convert the collaborative information into a binary format in the crafted prompts to guide the model in understanding recommendation tasks. 
% explores prompt strategies designed to enhance model performance through various instruction-tuning approaches, where prompts are crafted to guide the model in understanding recommendation tasks by predicting the next token based on designed instructions, often utilizing binary classification loss for specific outputs like "yes" or "no."
% kdd24_2 arxiv23_2 sigir24 tokenization
Another research direction explores different item identification to accurately generate potential items and avoid hallucination, especially in zero-shot settings.
TransRec~\cite{lin2024bridging}, an LLM-based recommender system uses multi-facet identifiers (ID, title, and attributes) to bridge the item and language spaces and further ensures accurate cold-start item recommendations through position-free constrained generation, using specific data structure (FM-index) to generate valid identifiers.
RecSysLLM~\cite{chu2023leveraging} regards the user/item attributes as entity tokens for identification and establishes an entity pool in a tree structure to facilitate the searching process in the zero-shot recommendation.
IDGenRec~\cite{tan2024idgenrec} uses an ID generator to create textual IDs for items and involves alternately training the ID generator and the base recommender with a specialized learning objective to encode items into concise, semantically rich textual IDs.

%% zhang2024notellm,zhang2024notellm,shen2024exploring,ma2024xrec
\smallskip\textit{\textbf{Fine-Tuning.}}
To move beyond instruction-based adaptation to recommendation contexts, fine-tuning-based methods adapt pre-trained LLMs to recommendation tasks through specifically designed losses (e.g., contrastive loss) or extra trainable parameters (e.g., soft prompts~\cite{yang2024item} and adaptors~\cite{ma2024xrec}) that enable the model to learn rich, recommendation-domain-specific patterns. The primary goal is to encode collaborative and semantic information more explicitly, allowing them to capture intricate user-item interactions and deliver accurate recommendations for new items and users with minimal or no prior interaction history.
% www24 arxiv24_1 arxiv24_3
For example, NoteLLM~\cite{zhang2024notellm} employs a generative-contrastive learning approach to integrate collaborative signals and also uses collaborative supervised fine-tuning loss to train the model on generating hashtags and categories. Following this, NoteLLM-2~\cite{zhang2024notellm} integrates contrastive learning and a late fusion mechanism for multimodal representation, while URLLM~\cite{shen2024exploring} leverages contrastive learning and domain-specific retrieved augmented generation to fine-tune LLMs for effective and domain-aligned recommendations.
% cikm23 arxiv24_4 
Adding extra modules for parameter-efficient fine-tuning is considered an alternative way to incorporate collaborative filtering knowledge from warm users/items. 
POD~\cite{li2023prompt} introduces a prompt distillation strategy distilling discrete prompts into continuous prompt vectors as extra parameters.
XRec~\cite{ma2024xrec} uses a simple adapter to bridge the gap between collaborative filtering signals and the LLM's semantic space.
In contrast, TALLRec~\cite{bao2023tallrec} and its follow-ups \cite{zhang2023collm,kim2024large,zhang2024text} employ a lightweight tuning method LoRA (Low-Rank Adaptation)~\cite{hu2022lora} to adapt LLMs efficiently.

\subsection{LLM as the Knowledge Enhancer}\label{sec:llm_as_enhancer}
The goal of cold-start recommendation tasks in specific scenarios is to use additional information to represent the preferences of cold instances as accurately as possible, and the world knowledge obtained by large language models through pre-training on a vast amount of corpus information can serve as a powerful knowledge base for warming up cold instances.

\subsubsection{LLM for Representation Enhancement}
Traditionally, the recommendation is mainly based on ID embeddings for users and items~\cite{he2020lightgcn,wang2019neural}, in which embeddings are randomly initialized and trained based on collaborative signals over historical interactions. However, due to the absence of available information for cold users/items, the ID embeddings of these instances hardly represent them accurately.
In this way, with the encoded world knowledge, LLMs can be adopted as the representation enhancer that (i) extends the original ID embeddings with other modality-aware representations, such as textual features and multimodal information, and (ii) models multi-domain knowledge with a unified LLM-based architecture. Based on the enriched user/item representations, the online recommender can provide more accurate user/item modeling.

%kim2024general,ren2024easyrec,hu2024enhancing
%tang2023one,gong2023unified,mysore2023large
\smallskip\textbf{\textit{Modality-Enhanced Representation}}. Commonly, recommendation models adopt ID embeddings that start from random initialization, which are only based on the model to aggregate information.
For representation enhancement with other modality knowledge, LLMs are always adopted as auxiliary encoders due to their powerful linguistic encoding ability~\cite{zhu2024collaborative,zheng2024adapting} or multimodal encoding ability by MLLMs~\cite{liu2024alignrec,liu2024multimodal}.
This type of model's key technical challenge is to effectively align the information of different modalities.
For example, Kim et.al~\cite{kim2024general} proposes a general item representation learning framework for cold-start content recommendation. This framework is not dependent on specific domains or datasets and can naturally integrate multimodal features through a Transformer-based architecture. %The model can be trained end-to-end without the need for categorical labels, directly leveraging user behavior data to learn item representations. 
EasyRec~\cite{ren2024easyrec} is a simple yet effective method that combines text-based semantic understanding and collaborative signals using a text-behavior alignment framework. It integrates contrastive learning and collaborative language model tuning to ensure an alignment between the text-enhanced semantic space and collaborative behavioral information. SAID~\cite{hu2024enhancing} utilizes a projection module to convert item IDs into embedding vectors and then leverages LLMs to explicitly learn embeddings that are semantically aligned with the textual descriptions of items. 
%In this way, SAID avoids the long text sequence issues encountered when directly using LLMs and reduces resource requirements in industrial scenarios.

%tang2023one,gong2023unified,mysore2023large
\smallskip\textbf{\textit{Domain-Enhanced Representation}}. Existing cross-domain recommendation systems typically require the design of complex model architectures to explore the relationships between two domains, making it difficult for models to scale to multiple domains and leverage more data. Moreover, existing recommendation systems use IDs to represent items, which carry fewer transferable signals in cross-domain scenarios, and user cross-domain behavior is also sparse, making it challenging to learn item relationships from different domains. 
Due to the architectural advantage of LLMs, which can unify encoding information in a single framework, LLMs are flexible enough to learn user/item representation in multiple domains simultaneously. For instance, LLM-REC~\cite{tang2023one} mixes user behavior across different domains and models user behavior using pre-trained language models, expecting to leverage the common knowledge encoded in pre-trained language models.
%to alleviate data sparsity and cold start problems. 
Gong et.al~\cite{gong2023unified} proposed a unified foundational model for search and recommendation, leveraging LLMs to extract domain-invariant text features, and fusing ID features, text features, and task-specific sparse features through aspect gating fusion to obtain representations for queries in search and cold items in recommendation. 
%KAR~\cite{mysore2023large} further proposes an open-world knowledge-enhanced recommendation framework that allows LLMs to acquire inferential knowledge about user preferences and factual knowledge about items. 
KAR~\cite{mysore2023large} generates inferential and factual knowledge by an open-world knowledge-enhanced recommendation framework, and then effectively transforms and compresses this knowledge into enhanced vectors through a hybrid expert adapter, making them compatible with recommendation tasks. These enhanced vectors can be directly used to enhance the performance of any recommendation model.

\subsubsection{LLM for Relation Augmentation}
%huang2024large,wang2024large
Another approach to leveraging LLMs as knowledge enhancers is through behavior augmentation. Specifically, the extensive world knowledge embedded in LLMs can be harnessed to analyze the potential preferences and interests of cold instances, thereby enhancing their potential behaviors. 
Since existing recommendation models predominantly rely on behavioral features for modeling~\cite{he2017neural,he2020lightgcn}, accurately performing this enhancement can substantially boost the modeling capabilities of current recommendation models for cold instances.

\smallskip\textbf{\textit{Behavior Simulation}}. Based on textual (or multimodal) information of user/item instances, LLMs can be adopted as the behavioral signal generator by analyzing the semantic similarity between pairs of users and items. The effectively generated behaviors can help to address the data sparsity issue of cold instances.
Specifically, ColdLLM~\cite{huang2024large} leverages a customized LLM simulator to mimic interactions (behavior patterns) between users and cold items, thereby directly transforming cold items into warm ones. In this way, cold and warm instances can be trained into a unified scheme with both simulated and real interactions. Similarly, Wang et.al~\cite{wang2024large} utilize LLMs to infer users' preferences for cold-start items based on the textual descriptions of their historical behaviors and the descriptions of new items, and then integrate these enhanced training signals into the learning of downstream recommendation models through auxiliary pairwise losses.

\smallskip\textbf{\textit{External Relation Supplement}}. Existing knowledge-based recommendation systems rely on limited metadata in knowledge relations from private resources, which typically item attributes and user interaction data, but they have limitations in dealing with cold start problems and data sparsity. LLMs can be used as discriminative intermediate enhancers, leveraging world knowledge and learned common-sense information to associate user-item interaction graphs with additional provided text-powered relation information, such as external text-rich knowledge graphs, to avoid the knowledge sparsity of cold instances.
CSRec~\cite{yang2024common} leverages common-sense knowledge from LLMs to construct a knowledge graph and combines it with existing knowledge-based recommendation methods. The CSRec framework effectively integrates common-sense knowledge from LLMs with metadata knowledge graphs through a common-sense knowledge graph and a knowledge fusion method.
CoLaKG~\cite{cui2024comprehending} leverages LLMs to enhance KGs by transforming graph data into textual inputs and generating semantic embeddings, which integrates global knowledge graph information to boost recommendation.

\section{CHALLENGES AND FUTURE OPPORTUNITIES}
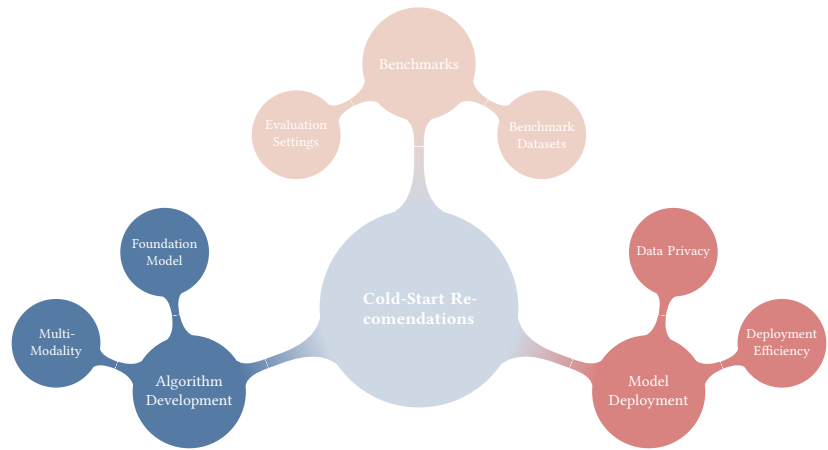
\begin{figure}[htbp]
\centering
\begin{tikzpicture}[scale=0.65, transform shape]
    \path[mindmap,concept color=treeroot, text=white, font=\bfseries]
    node[concept] {Cold-Start Recomendations}
    child[concept color=tree1,grow=-20]{
        node[concept] {Model Deployment}
        child[grow=20] {node[concept] {Deployment Efficiency}}
        child[grow=80] {node[concept] {Data Privacy}}
    }
    child[concept color=tree2,grow=200]{ 
        node[concept] {Algorithm Development}
        child[grow=160] {node[concept] {Multi-Modality}}
        child[grow=100] {node[concept] {Foundation Model}}
    }
    child[concept color=tree3,grow=90]{
        node[concept] {Benchmarks}
        child[grow=-30] {node[concept] {Benchmark Datasets}}
        child[grow=210] {node[concept] {Evaluation Settings}}} 
    ;
\end{tikzpicture}
\caption{Challenges and Future Opportunities of Cold-Start Recommendation.}
\label{fig:future}
\vspace{-3mm}
\end{figure}

In this section, we will discuss current challenges and future opportunities for cold-start recommendations. 
As shown in Figure~\ref{fig:future}, we will unfold the discussion with \textbf{Algorithm Development} (Sec.~\ref{sec:mm_csr} and Sec.~\ref{sec:fm_csr}), \textbf{Model Deployment} (Sec.~\ref{sec:eff_csr} and Sec.~\ref{sec:privacy_csr}), and \textbf{Benchmarks} (Sec.~\ref{sec:bench_csr}).

\subsection{Multi-Modal Cold-Start Recommendation}\label{sec:mm_csr}
Leveraging multi-modal information has emerged as a promising approach in modern recommendation systems \cite{zhou2023comprehensive, zou2024implicitave}, as it provides richer and more comprehensive representations of both items and user preferences through diverse modalities such as text, images, audio, and video. This capability is particularly valuable in addressing the cold-start problem by providing comprehensive user/item characteristics complementary to limited historical interactions. Despite the promising potential of multi-modal cold-start recommendations, several key challenges and opportunities remain for future research. 
An inherent challenge lies in effectively fusing and leveraging multi-modal information, as inappropriate utilization methods may introduce noise and degrade performance \cite{zhou2023comprehensive}. The prevailing issue of missing modalities in real-world applications poses another key challenge, as many existing multi-modal models assume all modality information is available during both training and inference \cite{wang2018lrmm, moscati2024multimodal}. Most existing recommendation datasets provide limited modalities, also constraining the application of state-of-the-art multi-modal methods to recommendation systems \cite{chen2024multi}. Additionally, a notable gap exists between user interest modeling and multi-modal embedding extraction, where pre-trained encoders are often directly utilized or optimized for content-oriented tasks rather than user preferences, resulting in a discrepancy between content understanding and personalization. Furthermore, existing approaches often fail to account for users' varying sensitivities to different modalities, as individuals may exhibit stronger preferences for certain modal aspects while remaining indifferent to others \cite{ye2024few}.
To address these challenges, we propose several potential research directions: (1) developing effective and efficient modality fusion methods that capture complementary information while remaining robust to noise and missing data; (2) constructing modality-rich recommendation datasets that incorporate diverse modalities beyond textual and visual information; (3) bridging the gap between multi-modal content and user interest modeling through personalized multi-modal embedding techniques and end-to-end architectures that jointly optimize content understanding with user preferences; and (4) designing adaptive personalization frameworks that dynamically adjust the importance of different modalities based on individual user preferences and contexts.

\subsection{Recommendation Foundation Models}\label{sec:fm_csr}
The ongoing transformation within the field of Natural Language Processing (NLP) has been powered by the emergence of foundation models—comprehensive, pre-trained models that can be easily adapted to a variety of downstream tasks. These models, such as large language models (LLMs), have demonstrated remarkable capabilities in tasks ranging from text classification to dialogue generation, significantly reducing the need for task-specific training from scratch. Inspired by these developments, the recommendation domain now faces a similar opportunity: to leverage analogous “foundation models” for a wide range of cold-start tasks, thereby enhancing adaptability, efficiency, and scalability.
As discussed in Section \ref{llm}, existing research has shown that LLMs can effectively address the cold-start issue for single tasks or within single domains. However, these solutions often require continuous long-term re-adaptation when encountering new recommendation tasks or domains, leading to substantial time and computational overhead. 
Future research directions involve the design and implementation of these recommendation foundation models, as well as algorithms to quickly adapt large foundation models to different CSR subtasks. Such models would not only enable multi-task recommendation capabilities, as exemplified by P5~\cite{geng2022recommendation}, but also dynamically adjust their recommendations across diverse domains.

\subsection{Efficiency in Cold-Start Recommendations}\label{sec:eff_csr}
Though current approaches for solving cold-start recommendations have shown promising results, they primarily focus on offline setups and face significant challenges when deployed on large-scale real-world systems due to the high latency and resource cost it will introduce~\cite{wang2024fresh}. For example, current LLM-based CSR approaches often suffer from significant computational overhead during both training and inference, posing challenges for real-time deployment. There is a great need for solutions capable of handling industrial-scale RecSys while meeting the query-per-second requirements of real-time applications.
Future research could prioritize developing lightweight and scalable models and exploring hybrid strategies that combine content-based and collaborative filtering techniques for more robust solutions. Real-time learning mechanisms are particularly critical, as they can swiftly incorporate early signals from cold-start items or users into the feedback loop, allowing for rapid adaptation to dynamic environments. Specifically, to support the applications of LLMs in cold-start recommendations, hybrid solutions such as hierarchical planning could be explored. For instance, LLMs could be utilized for high-level planning, such as cold-start content or user cluster selection, while traditional recommendation models focus on real-time, low-level item recommendations to ensure both efficiency and accuracy \cite{wang2024llms}.

\subsection{Data Privacy in Cold-Start Recommendations}\label{sec:privacy_csr}
Data privacy has long been a challenge for recommendation systems~\cite{huang2019privacy,himeur2022latest}, and the reliance on user information beyond interactions makes the challenges of cold start recommendations even more pronounced. For example, there are some listed issues:
(C-i) \textbf{Dependency on side information}. Building user profiles is central to recommendation systems, but it can involve the use of sensitive information. Striking a balance between utilizing effective user-profiles and protecting user privacy is a challenge in the design of cold-start models. (C-ii) \textbf{Domain information privacy.} Privacy preservation in cross-domain recommendation systems is more challenging, especially in cold start scenarios where data sparsity is a significant issue. The need to encourage collaboration between different domains for data can lead to privacy breaches if not managed carefully.
To address these issues, there are some technologies that can be promising: (T-i) \textbf{Privacy calculation}. Differential privacy calculation can protect user privacy information by adding noise to the data, making it impossible for the platforms and attackers to infer information about any specific individual through analysis of the results~\cite{wang2023improved,shin2018privacy}. In addition to that, homomorphic encryption allows computations to be performed directly on encrypted data without the need for decryption, thus protecting the privacy of the data~\cite{kim2018efficient,jumonji2021privacy}.
(T-ii) \textbf{Federated learning}. Federated learning (FL) offers several advantages for privacy protection in the context of cold start issues~\cite{yang2020federated,javeed2023federated}. FL allows data, like side information, to remain local and not uploaded to a central server, thereby protecting user privacy. FL enables decentralized model training, where multiple parties can collaboratively train a model without sharing raw data, which is helpful for cross-domain cold-start recommendations~\cite{meihan2022fedcdr,tian2024privacy}. Further, in FL, only model updates (such as gradients) are transmitted to a central server for aggregation, not raw data. This further reduces the risk of privacy leaks in cold start issues.

\subsection{Benchmark and Unified Evaluation}\label{sec:bench_csr}
Research on cold start recommendation systems has made significant progress.
However, evaluations for existing cold start recommendation systems are currently diverse and inconsistent. 
Thus, it would be promising and meaningful to develop a unified and fair evaluation benchmark for the community.
Specifically, there are multiple problem settings for cold-start recommendations, such as strict cold-start, non-strict cold-start, and long-tail cold-start, and each setting needs to be evaluated fairly. This raises the following three main issues:
(i) \textbf{Different benchmark datasets}.
Different papers rarely overlap in the datasets they use, making it difficult for researchers to conduct unified comparisons across several related datasets. A promising approach for the future would be to encourage the community to focus on a selection of high-quality datasets for comparative experiments. Further, only part of the papers adopts datasets with real cold users/items in industrial platforms. We encourage the community to release these datasets for open research, which is very meaningful for the development of the community. 
(ii) \textbf{Different evaluation settings}. The key problem in this part is the setting of cold users/items. Specifically, how to define and synthesize cold users/items in the experiments? 
In the experiments of many cold-start works, cold users/items are often obtained by varied approaches (time range, number of interactions, and random selection) with vague definitions of CSR settings, and there is still a lack of unified rules.
(iii) \textbf{Open-source evaluation frameworks}. There are some practical open-source projects for recommendation experiments, such as Recbole~\cite{zhao2021recbole,zhao2022recbole}, Elliot~\cite{anelli2021elliot}, and BARS~\cite{zhu2022bars}. However, all of them are designed for general recommendation evaluations. A unified open-source evaluation codebase for cold-start recommendations with fair evaluation settings and metrics will be very helpful for the research community.

\section{CONCLUSION}
In this paper, we provide a comprehensive review of cold-start recommendations, with a roadmap from content features, graph relations, and domain information, toward world knowledge from large language models. Specifically, we first formally define different research questions in the field of cold-start recommendations. Then, we systematically review cold-strat recommendations. In each part, we provide overall insights behind related works and list some representative works for readers to better understand. Furthermore, we rethink some of the challenges of cold-start recommendations and summarize some meaningful future directions. Related resources are organized in the Github (\url{https://github.com/YuanchenBei/Awesome-Cold-Start-Recommendation}) for the CSR research and industrial community.

% \begin{table}
%   \caption{Frequency of Special Characters}
%   \label{tab:freq}
%   \begin{tabular}{ccl}
%     \toprule
%     Non-English or Math&Frequency&Comments\\
%     \midrule
%     \O & 1 in 1,000& For Swedish names\\
%     $\pi$ & 1 in 5& Common in math\\
%     \$ & 4 in 5 & Used in business\\
%     $\Psi^2_1$ & 1 in 40,000& Unexplained usage\\
%   \bottomrule
% \end{tabular}
% \end{table}

% \begin{table*}
%   \caption{Some Typical Commands}
%   \label{tab:commands}
%   \begin{tabular}{ccl}
%     \toprule
%     Command &A Number & Comments\\
%     \midrule
%     \texttt{{\char'134}author} & 100& Author \\
%     \texttt{{\char'134}table}& 300 & For tables\\
%     \texttt{{\char'134}table*}& 400& For wider tables\\
%     \bottomrule
%   \end{tabular}
% \end{table*}

%%
%% The acknowledgments section is defined using the "acks" environment
%% (and NOT an unnumbered section). This ensures the proper
%% identification of the section in the article metadata, and the
%% consistent spelling of the heading.
% \begin{acks}
% To Robert, for the bagels and explaining CMYK and color spaces.
% \end{acks}

%%
%% The next two lines define the bibliography style to be used, and
%% the bibliography file.
\bibliographystyle{ACM-Reference-Format}
\bibliography{main/sample-base}

%\clearpage
%\input{main/9_appendix}

%%
%% If your work has an appendix, this is the place to put it.
%\appendix

\end{document}